\documentclass[a4paper]{elsart}
\pdfoutput=1
\usepackage[latin1]{inputenc}
\usepackage{caption}
\usepackage{subcaption}
\usepackage{graphicx}
\usepackage{epsfig}
\usepackage{wrapfig,rotating}
\usepackage{amssymb,amsmath,array}
\usepackage{epstopdf}
\usepackage{hyperref}
\hypersetup{
  colorlinks=true,
  linkcolor=red,
  citecolor=red,
  urlcolor=blue
}

\DeclareCaptionLabelFormat{andtable}{#1˜#2 \& \tablename˜\thetable}
\begin{document}
\begin{frontmatter}
\title{Forward tracking at the next \boldmath{$e^+e^-$} collider \\ part II: experimental challenges and detector design}
\author[desy]{S. Aplin}
\author[valencia]{M. Boronat} 
\author[cern]{D. Dannheim}
\author[cantabria]{J. Duarte}
\author[desy]{F. Gaede}
\author[cantabria]{A. Ruiz-Jimeno}
\author[cern]{A. Sailer}
\author[hephy]{M. Valentan}
\author[cantabria]{I. Vila}
\author[valencia]{M. Vos\corauthref{cor}}
\corauth[cor]{Corresponding author. E-mail: marcel.vos@ific.uv.es}
\address[desy]{Deutsche Elektronen Synchrotron, Hamburg, Germany}
\address[valencia]{IFIC (U. Valencia/CSIC), Apd. Correos 22085, E-46071 Valencia, Spain}
\address[cern]{CERN, CH-1211 Geneva 23, Switzerland}
\address[cantabria]{IFCA (U. Cantabria/CSIC), A. de los Castros s/n, E-39005 Santander, Spain }
\address[hephy]{HEPHY, Nikolsdorfer Gasse 18, 1050 Vienna, Austria}

%


\begin{abstract}
We present the second in a series of studies into the forward tracking system 
for a future linear $ e^+ e^- $ collider with a center-of-mass energy in the 
range from 250~GeV to 3~TeV. In this note a number of specific challenges are
investigated, that have caused a degradation of the tracking and vertexing 
performance in the forward region in previous experiments. We perform a 
quantitative analysis of the dependence of the tracking performance on 
detector design parameters and identify several ways to mitigate the 
performance loss for charged particles emitted at shallow angle.
\end{abstract}
\begin{keyword}
Forward tracking, future linear $e^+ e^- $ collider, ILC, CLIC
\end{keyword}
\end{frontmatter}

\section{Introduction}

A high-luminosity, high-energy, linear $e^+ e^-$ collider yields excellent 
opportunities for precision tests of the Standard Model of particle physics. 
The combination of precisely calculable electroweak production and 
strict control of the initial
state with the relatively benign experimental environment and 
state-of-the-art detector systems allow for a characterization
of Standard Model and new physics processes with a precision 
that goes well beyond what can be achieved at hadron colliders.

Two projects exist that pursue the creation of a linear electron-positron 
collider (referred to as Linear Colliders or LC in the remainder of this 
note):
\begin{itemize}
\item The proposal for an International Linear Collider 
(ILC~\cite{BrauJames:2007aa}) is based on existing super-conducting 
Radio-Frequency (RF) cavity technology. In the baseline design the ILC is 
envisaged to reach a center-of-mass energy of 500 GeV. Early stages
of the physics programme are likely to involve running at a 
center-of-mass energy of 250--350 GeV to study the properties 
and couplings of the Higgs boson and to characterize the 
production threshold for $t \bar{t}$ pair production.
The possibility to upgrade the ILC to a maximum of $ \sqrt{s} = $ 1~TeV is a 
crucial requirement to the design. 
\item To reach larger center-of-mass energies, the accelerating 
gradient obtained in the previous scheme is insufficient. 
The Compact Linear Collider (CLIC~\cite{cliccdr1}) aims to open up 
the energy regime up to several TeV using a novel technology, where a 
drive beam is used to provide power to the room temperature RF cavities 
of the main Linac.
\end{itemize}

The physics case for a linear $e^+e^-$ machine has been made in great detail
in References~\cite{Brau:2012hv,Djouadi:2007ik,snowmass1,snowmass2,snowmass3,snowmass4,tesla}. The specific case of a multi-TeV $ e^+ e^- $ collider 
is discussed in References~\cite{cliccdr2,cliccdr3,clicphysics}. 
Recently, the focus 
has naturally shifted to a precise determination of the properties of the 
boson~\cite{:2012gk,:2012gu} discovered at the Large Hadron Collider (LHC). 
However, a linear collider that covers the energy regime from several hundreds
of GeV to several TeV offers a much broader programme of Standard 
Model measurements and searches for new phenomena. The interplay of 
the broader LC programme with respect to the LHC is studied in 
Reference~\cite{Brau:2012hv,weiglein}.

A very active programme exists aimed at the development of detectors for 
a linear collider~\cite{ilcdetectors}. Innovative approaches towards 
calorimetry, tracking and vertex detectors are pursued in detector 
\mbox{R \& D} collaborations. Two detector concept groups have prepared 
complete detector designs~\cite{ildloi,sidloi}. The same two detector 
concepts have been adapted to the CLIC environment~\cite{cliccdr2}. 

In a previous article~\cite{forwardnote} we discussed the relevance
of the detector performance at small polar angle for a number of key LC physics 
analyses. It is found that increasingly abundant many-fermion final states 
are often not contained in the central detector. We have moreover 
identified several potentially important processes, 
such as di-boson production, t-channel production of new particles, 
or Higgs boson production through vector boson 
fusion, that exhibit a strong preference for the forward region.
This study established that, as the center-of-mass 
energy enters the 0.5-3~TeV regime, the track reconstruction performance
 in the polar angle\footnote{Our notation corresponds to the usual cylindrical coordinates, where the $ z$-axis coincides with the beam line and the field lines of the solenoidal magnetic field, $ r $ denotes the radial distance from the beam line in the plane perpendicular to $ z$ and $ \phi $ is the azimuthal angle. We also use the polar angle $\theta$ that varies from $ \theta =$ 0$^o$ when the vector is aligned with the beam line, to $\theta =$ 90$^o$ when it is perpendicular. } region from 5$^\circ < \theta < $ 30$^\circ $, 
corresponding to \mbox{87 $ \rm ~mrad < \theta < $ 520 $ \rm ~mrad $}, 
or 0.87  $ < \cos \theta < $ 0.996, acquires a much greater relevance
than at previous $e^+e^-$ colliders. 

Reconstruction of charged particles in the polar angle 
range from \mbox{ 5$^\circ < \theta < $ 30$ ^\circ$} faces a number of 
specific challenges. The forward tracker in many experiments 
has not performed as well as the central tracker in key aspects 
like the momentum resolution, vertex reconstruction performance 
or the material budget. The aim of this paper is to quantify 
these effects and explore ways to mitigate their impact on the 
overall physics output of the experiment.

This paper is organized as follows. 
Section~\ref{sec:detectorconcepts} provides a brief introduction 
of the LC detector concepts. The software and tools used to 
perform the simulation studies are briefly 
outlined in Section~\ref{sec:software}.
In the subsequent sections we provide quantitative results on the 
following aspects of the tracker performance:
\begin{itemize}
\item Section~\ref{sec:backgrounds}: the contribution of beam-induced backgrounds to the hit density.
\item Section~\ref{sec:momentum}: the momentum measurement for charged 
particles.
\item Section~\ref{sec:vertex}: the reconstruction of the primary and 
secondary production vertices of charged particles.
\item Section~\ref{sec:pattern}: the pattern recognition capabilities of the 
detector.
\end{itemize}
In Section~\ref{sec:discussion} we summarize the findings. In an appendix a
number of technicalities of forward tracking are discussed.

\section{Detector concepts}
\label{sec:detectorconcepts}

Most collider experiments adopt a cylindrically symmetric geometry 
where the magnetic field in the 
tracking volume is provided by a large solenoid. This is the case for
all ILC and CLIC detector concepts proposed until today.
Many of the challenges of the forward tracking region are closely related to 
this geometry and are relatively insensitive to the details of the
detector design. In many cases we can expose the dependence of the performance
on detector design parameter by studying a generic simplified geometry.

To provide more quantitative and realistic estimates of the tracking 
performance, however, one needs to fill in the details of the detector 
design. In this note we consider a detailed detector concept 
developed for the ILC~\cite{ildloi} and later 
adapted to the CLIC environment~\cite{cliccdr2}.

\begin{figure}[h]
\centering
\includegraphics[width=\columnwidth]{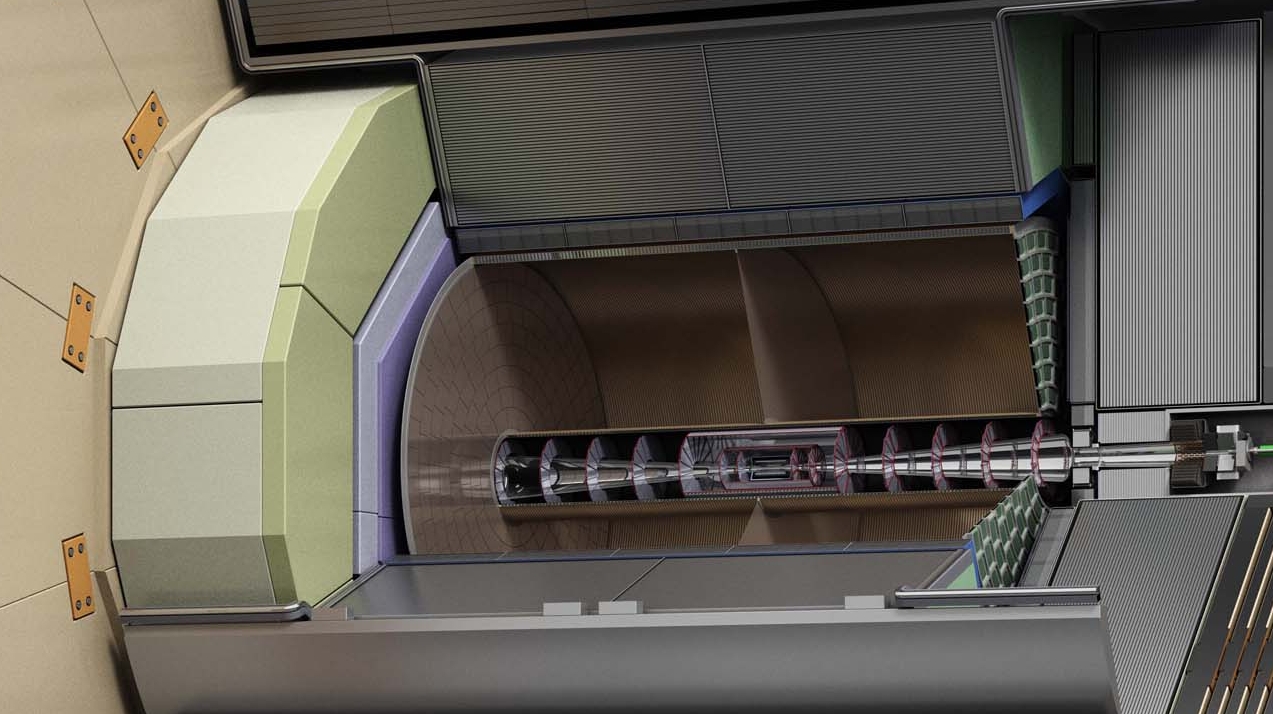}
\caption{A lateral view of the innermost part of the ILD detector. Figure prepared for the ILD Detailed Baseline Design~\cite{ilddbd}.}
\label{fig:quadrant}
\end{figure}

A lateral view of a quadrant of the ILD detector is shown in 
Figure~\ref{fig:quadrant}. A brief description of the most important
features of the ILD and SiD~\cite{sidloi} detector concept follows.

Both concepts are based on the particle flow concept that combines
the energy measurements of different sub-systems 
-- the tracker information for charged particles, the electromagnetic 
calorimeter for photons, and the hadronic calorimeter 
for neutral hadrons -- to achieve an unprecedented jet energy resolution. 
The inner layers of the calorimeter system are formed by a highly granular
Si-W calorimeter. In the outer layers iron may be used as absorber material 
(but running at a center-of-mass energy of one TeV and beyond may require
tungsten also for the hadronic calorimeter to achieve a compact system 
that can contain hadronic showers of highly energetic particles). 
The granularity is also reduced to the level that relatively large cells
(made of scintillating material) can yield an accurate measurement 
of the particle multiplicity. The calorimeter system is inserted in
a superconducting solenoid that provides an approximately axial
magnetic field with a strength of 3.5 Tesla for ILD (4 Tesla in 
the CLIC variant of the detector concept) and 5 Tesla for SiD.

Both concepts envisage a conical beam pipe that limits the acceptance
of the tracking system to approximately $5^\circ < \theta < 175^\circ$. 
In the Silicon Detector concept the entire tracking 
volume is instrumented with silicon-based devices. A compact pixel detector 
close to the interaction point acts as a vertex detector. 
The 5 cylindrical layers in the {\em barrel} part of the detector are
complemented by {\em end-cap} disks to provide nearly uniform 
five-point coverage down to a polar 
angle of 12.5$^\circ$. The vertex detector is surrounded by 5 barrel 
layers and 4 disks equipped silicon micro-strip detectors.


\begin{figure}[h]
\centering
\includegraphics[width=0.49\columnwidth]{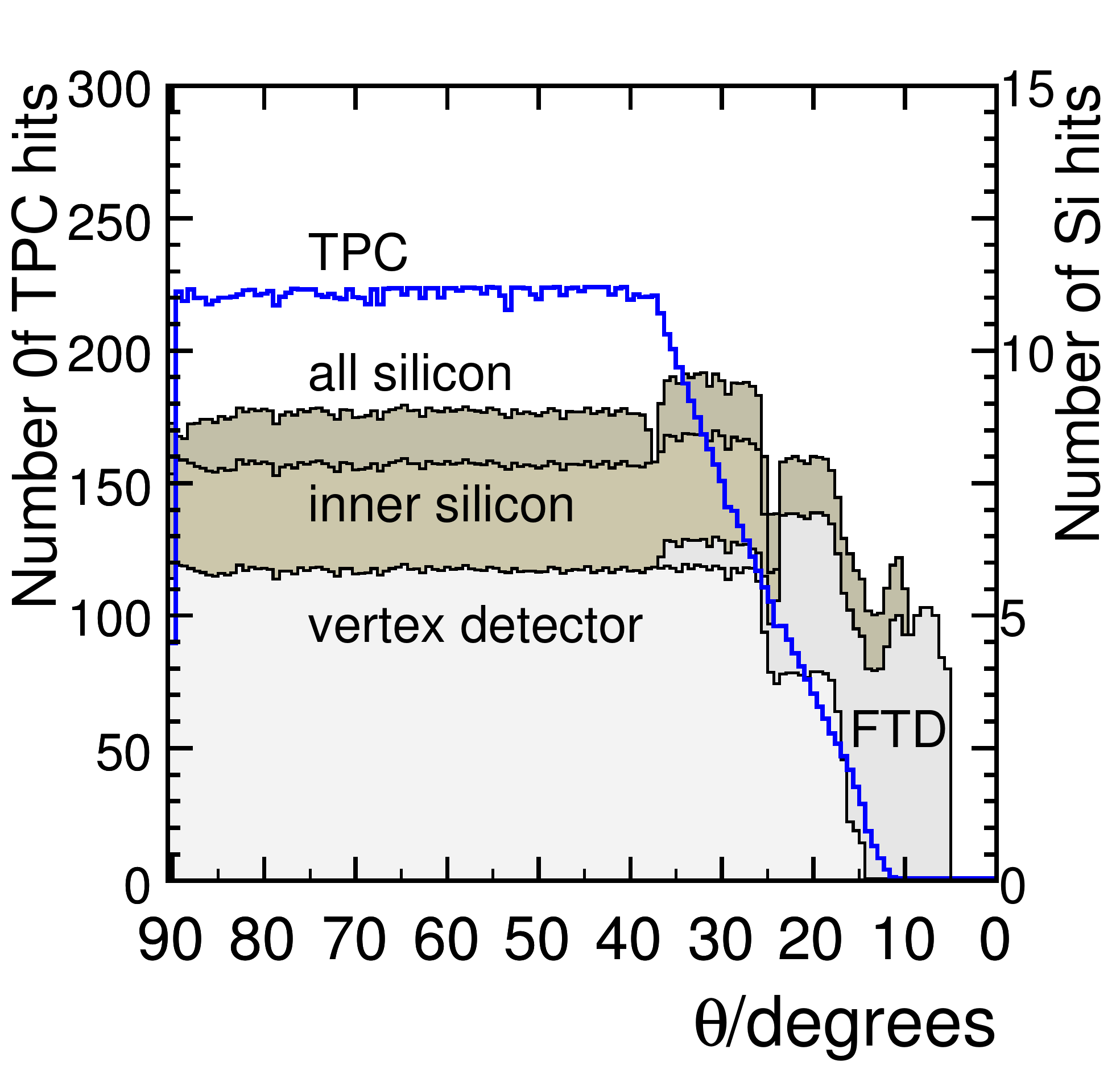}
\includegraphics[width=0.49\columnwidth]{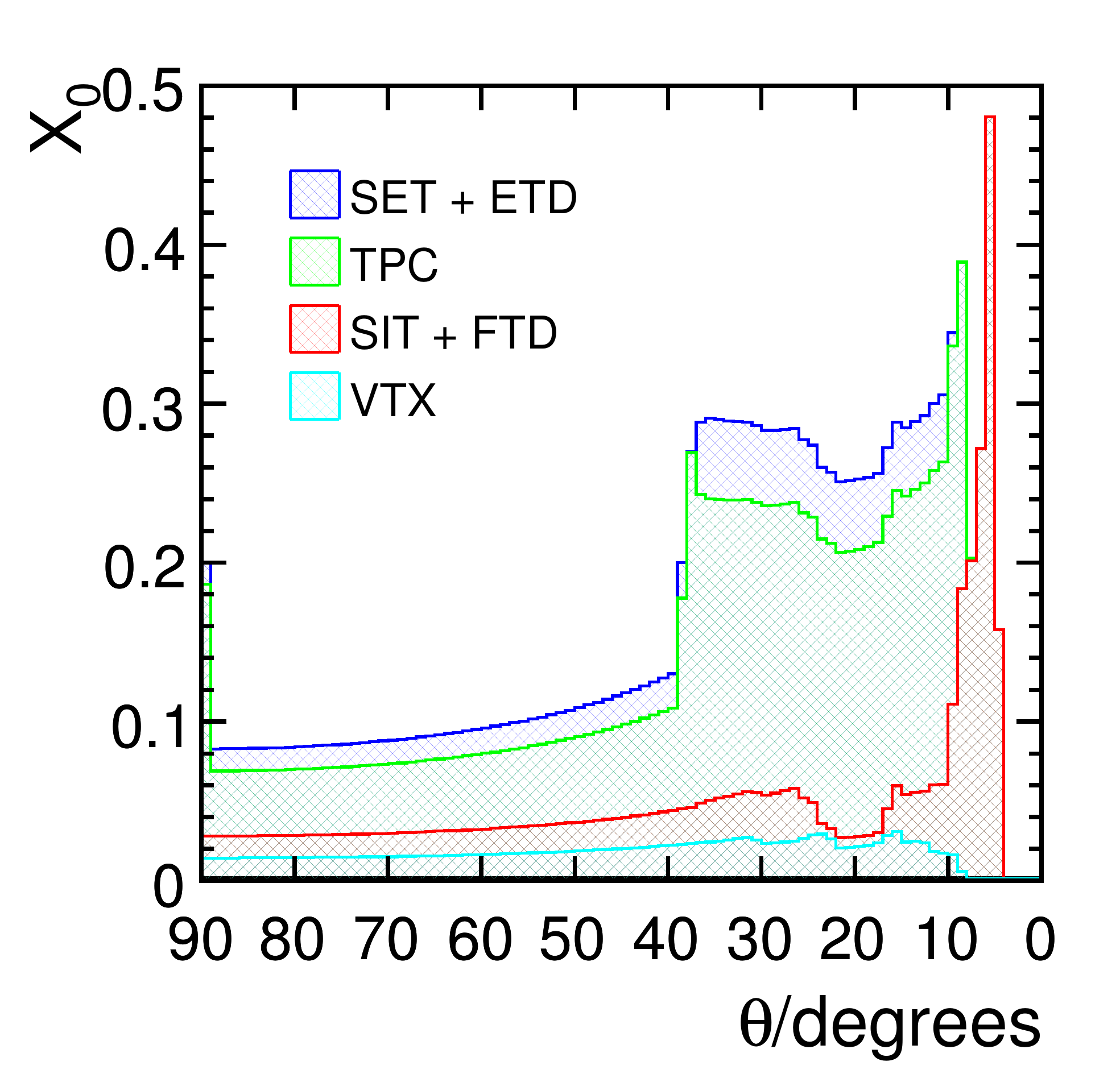}
\caption{Leftmost panel: the number of hits attached to 100 GeV muons in the gaseous and solid tracking subsystems versus polar angle $ \theta $ in the ILD design for the ILC. The number of hits in the TPC is referred to the leftmost axis. The number of hits in the silicon detectors (including the vertex detector) is indicated with different shadings of grey. Rightmost panel: the detector material in radiation lengths versus polar angle. The beam pipe is accounted for in the SIT+FTD material. Both figures are reproduced from Reference~\cite{ildloi}.}
\label{fig:trackingperf_coverage}
\end{figure}

The ILD~\cite{ildloi} detector relies on a large Time Projection Chamber 
for reconstruction of charged tracks. The vertex detector and innermost 
tracker layers are based on silicon pixel and micro-strip detectors, 
respectively. 
The choice of a smaller magnetic field (3.5 T instead of the 5 Tesla in SiD) 
drives the size of the detector. The outer envelope of the TPC is equipped 
with additional silicon micro-strip detectors. The polar angle coverage of 
the different subsystems of the ILD tracker is indicated in 
figure~\ref{fig:trackingperf_coverage}, where the number of hits attached to 
100 GeV muons is shown as a function of polar angle. The Time Projection 
Chamber provides full coverage down to a polar angle of approximately 
37$^\circ$. Below that angle the number of read-out rings gradually decreases. 
The last TPC ring corresponds to a polar angle of just over 10$^\circ$. 
The central innermost tracking system, consisting of the 3 $\times$ 2
layers of the barrel vertex detector and the two layers of the Silicon 
Internal Tracker (SIT), provides eight 
precise measurements down to 26$^\circ$. The innermost and middle double 
layer of the vertex detector extend out to approximately 16$^\circ$. The 
seven Forward Tracking Disks (FTD) provide up to five measurements for 
tracks at small polar angle. The Silicon External Tracker (SET) and Endcap 
Tracking Disks (ETD) provide a precise space point with large lever arm down 
to approximately 10$^\circ$. 

The ILD tracking detectors are extremely 
transparent. A thin Berylium beam pipe contributes 0.07\% of a radiation
length ($X_0$). The material budget of the vertex detector is
0.16\% $X_0$ per double ladder, each innermost
Forward Tracking Disks contributes only 0.12\% $X_0$, and the layers equipped
with $\mu$-strip detectors represent 0.65\% $X_0$/layer. 
Figure~\ref{fig:trackingperf_coverage} clearly indicates how the material in
the TPC is concentrated in the end-plate.

The design for the CLIC tracker elements is modified in several ways to 
cope with the increased in background levels~\cite{lcd-note-2011-031}. 
The most important modification from the point of view of the overall 
detector performance is the increase in the radius of the innermost 
vertex detector layer ($r=$ 2.6 or 3.1 cm instead of the approximately 1.5 cm in 
both concepts for the ILC). Of particular importance for the forward 
tracking region is the end-cap system of the pixel vertex detector, that is
added to the CLIC-ILD design to recover the forward coverage lost
due to the increase of the inner radius of the barrel system. The redesign
of the beam pipe leads to a slight loss of acceptance for very shallow tracks.

We have chosen to show results for the ILD concept. However, at the qualitative
level all conclusions apply also to the SiD tracker.

\section{Track reconstruction}
\label{sec:software}

An extensive set of tools is available to study the challenges of forward 
tracking at a quantitative level. We used two fast simulation packages,
the Linear Collider detector toy, or LiCToy~\cite{lictoytheprogram}, 
and a custom setup based on the CMS track fitter. 
LiCToy uses an idealised helix model, with 
multiple Coulomb scattering, to simulate the particle trajectory in a 
solenoidal field, while the second package propagates the trajectories 
of charged particles through an arbitrary magnetic field map. 
A simplified detector geometry is constructed based on simple geometric
shapes (cylinders and disks). The hit positions are smeared with a Gaussian
or uniform distribution.
In both packages the track fit is performed using a Kalman 
filter that takes into account interactions with the detector material.

The LC detector concepts have developed ({\em full}) simulation suites, 
relying on GEANT4~\cite{geant4} to simulate the interactions of particles 
with a detailed detector model. We use the chain consisting of
Mokka~\cite{lc-tool-2003-010} for the interactions with the detector,
the Marlin~\cite{Gaede:2006pj} package for reconstruction, and 
LCIO~\cite{Aplin:2012kj} to persistify objects.
The LDCTracking package involves a track fit taking into 
account material effects. Recently, a completely new tracking software 
toolkit~\cite{arlington} was adopted by ILD, including a pattern recognition 
algorithm for the forward direction based on cellular 
automatons~\cite{Glattauer:2012rh}.

The fast simulation tools were validated by comparing their 
results on a number of simple geometries. A further cross check was 
performed against the LCDTRK~\cite{lcdtrk} package, that is based 
on an analytical calculation of the covariance 
matrices~\cite{Billoir:1983mz}. All codes are found to agree within errors
when run on the same benchmark geometry.

\subsection{Forward track reconstruction}

In this paper we discuss a number of challenges of tracking in the forward 
region that are either not present or less severe in the central detector. 
Often, the forward region is considered even more special, so much so
that it would require dedicated track reconstruction algorithms. 
In the following we argue, by means of an example, that general-purpose 
3D trajectory propagation algorithms are adequate for the entire detector.

\begin{figure}
\begin{center}
\includegraphics[width=0.8\columnwidth]{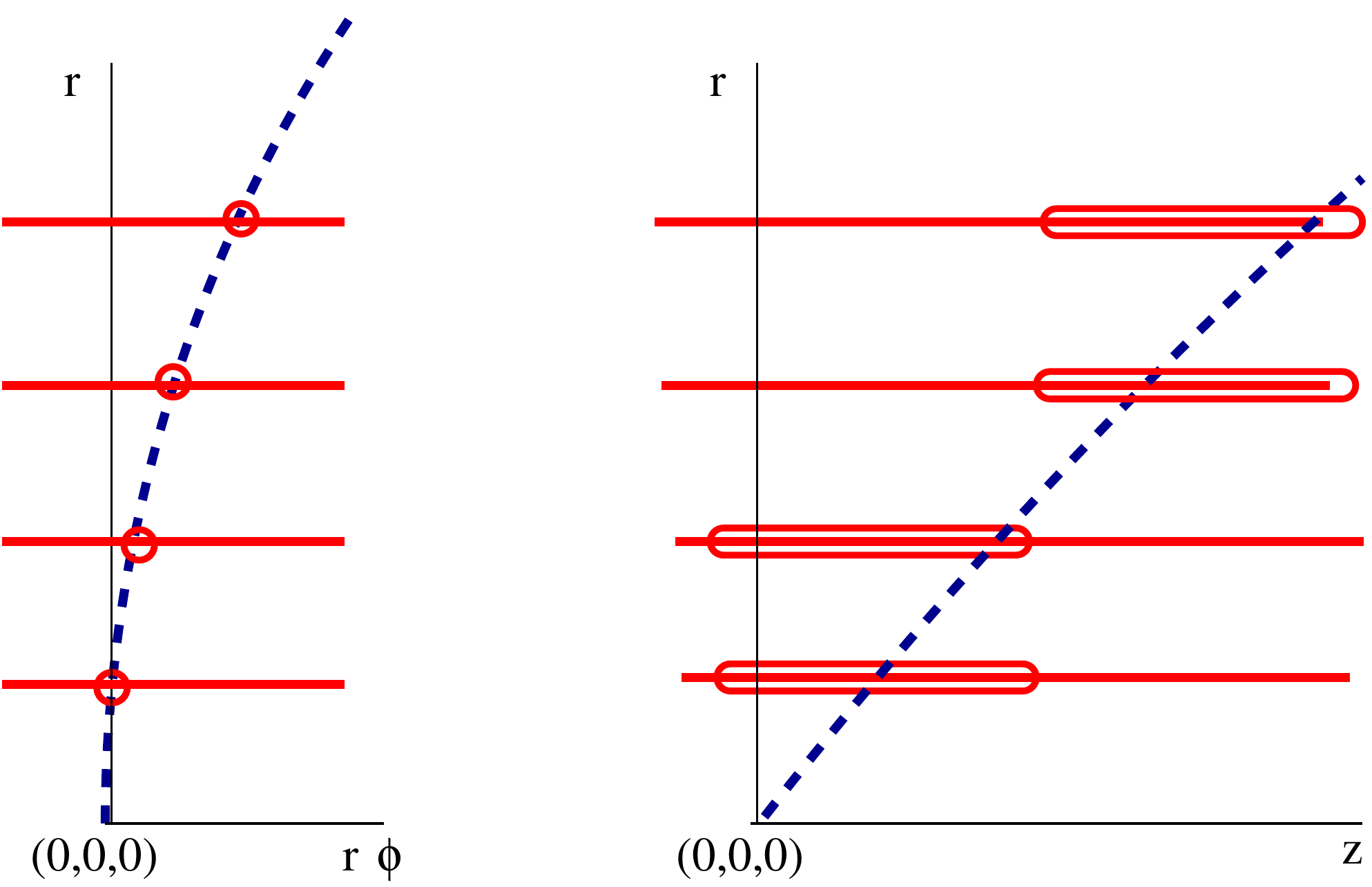} \\
\vskip 1cm
\includegraphics[width=0.8\columnwidth]{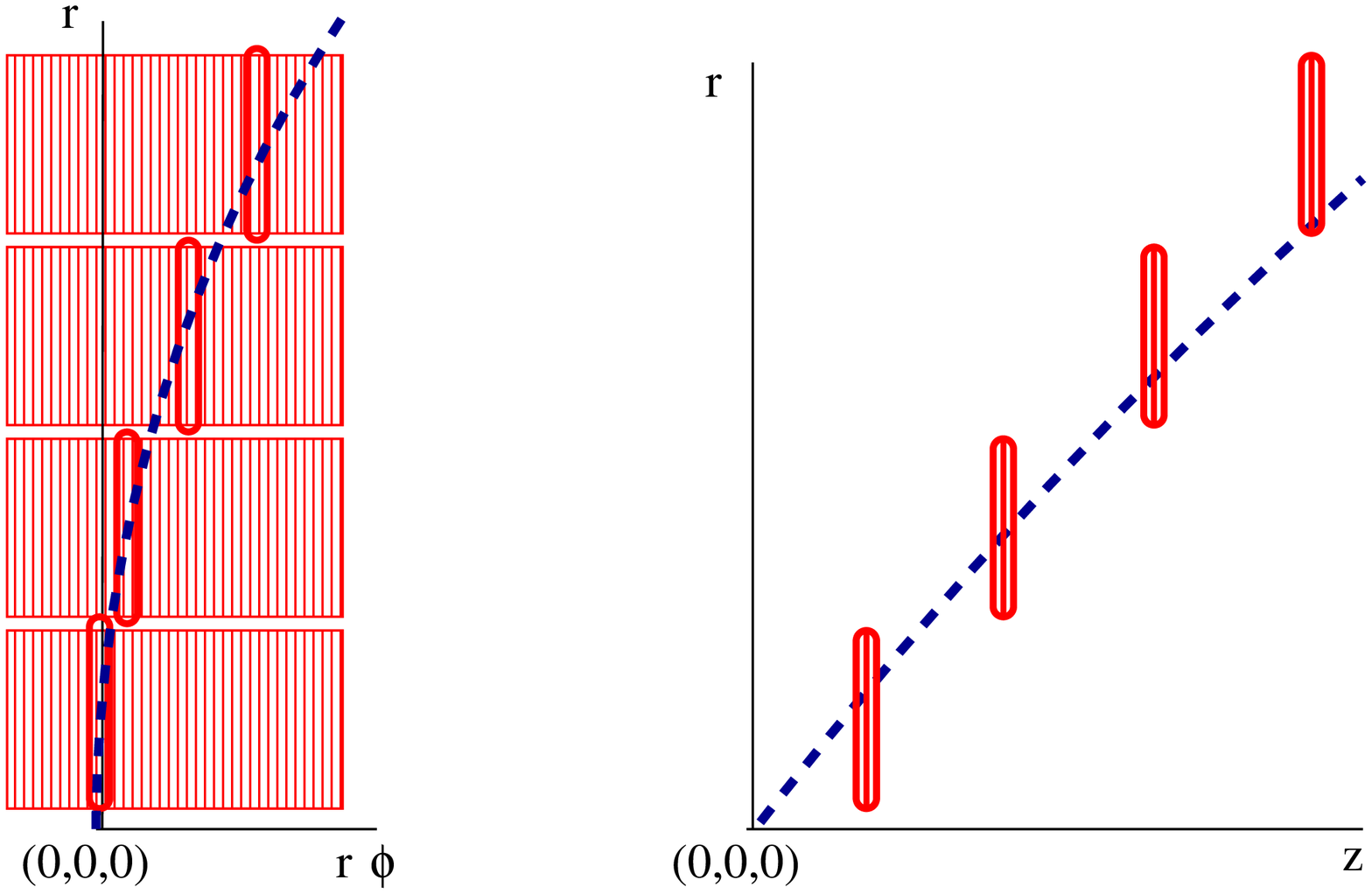}
\caption{The projections of a helix trajectory on the $ r \phi $ and $ rz$ planes. The hits on solid state detectors along the trajectory are indicated for two situations: the uppermost drawings represent a cylindrical {\em barrel} detector geometry and the lowermost drawings correspond to the {\em endcap} of the tracker.}
\label{fig:trackfitviewsrphi}
\end{center}
\end{figure}

The trajectory of a charged particle in a uniform magnetic field is described 
by a helix. Five parameters are sufficient to describe the whole trajectory. 
For analytical calculations it is useful to break the helical trajectory down 
into two projections. In the $r-\phi$ view our ideal particle describes a 
circle. In the $ r-z $ view the helical trajectory yields a sine wave, often 
approximated by a straight line for high momentum tracks. In 
Figure~\ref{fig:trackfitviewsrphi} we show these two projections for a 
particle leaves the interaction point (0,0,0) at a polar angle of 45$^o$. 
In the uppermost Figure the track leaves hits on horizontal (barrel) layers
with a small error in $r \phi$ and a weak constraint on $z$. The second
set of figures in the lowermost panel corresponds to the same particle
traversing the {\em end-cap}. 

In the past, limitations of CPU time and other resources have led experiments 
to use the analytical solution for an ideal particle in the $r\phi$ and $rz$ 
projections for track finding or even the final track fit. In the projections
barrel and end-cap provide a different set of constraints on the trajectory.
In the $r \phi$ projection the trajectory, and in particular the 
momentum, is less constrained by the {\em end-cap} detector. 

In the last decades, however, a sophisticated machinery has been developed 
to deal with the fully general case. Numerical 3D trajectory propagation 
allows to remove the evident limitations of the analytical
approach, by taking into account energy loss and multiple scattering 
in detector material and by performing the propagation with a realistic 
magnetic field. All modern experiments use this much more general
approach in the final track fit. Even in areas, such as pattern recognition 
or the trigger, where one used to resort to a simpler treatment, the 3D 
formalism is rapidly taking over. It seems safe, therefore, to assume the 
ILC or CLIC experiments will have access to the general 3D treatment at 
all stages of track reconstruction, including track finding. 

\begin{figure}
\begin{center}
\includegraphics[width=0.48\columnwidth]{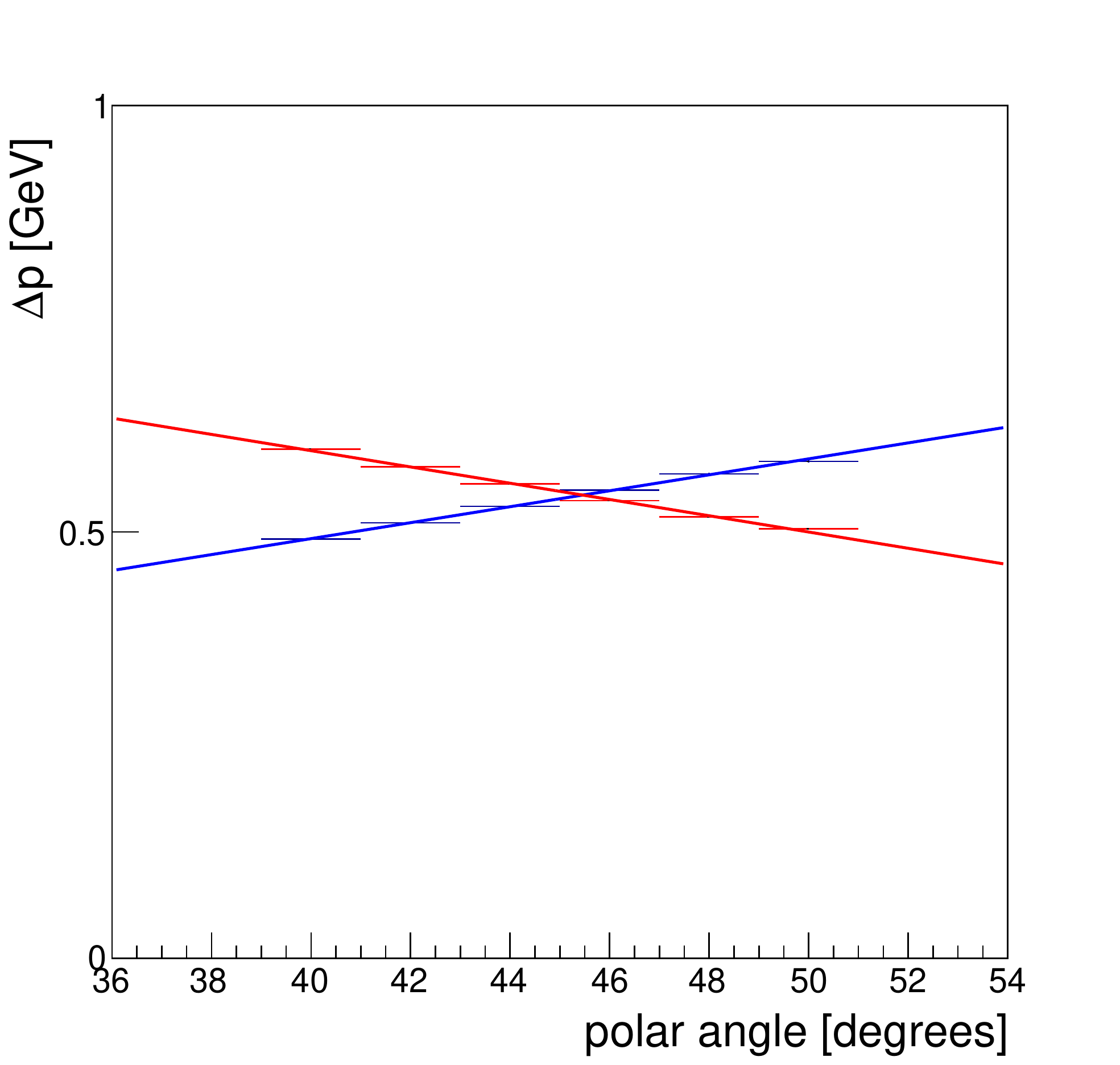} 
\includegraphics[width=0.48\columnwidth]{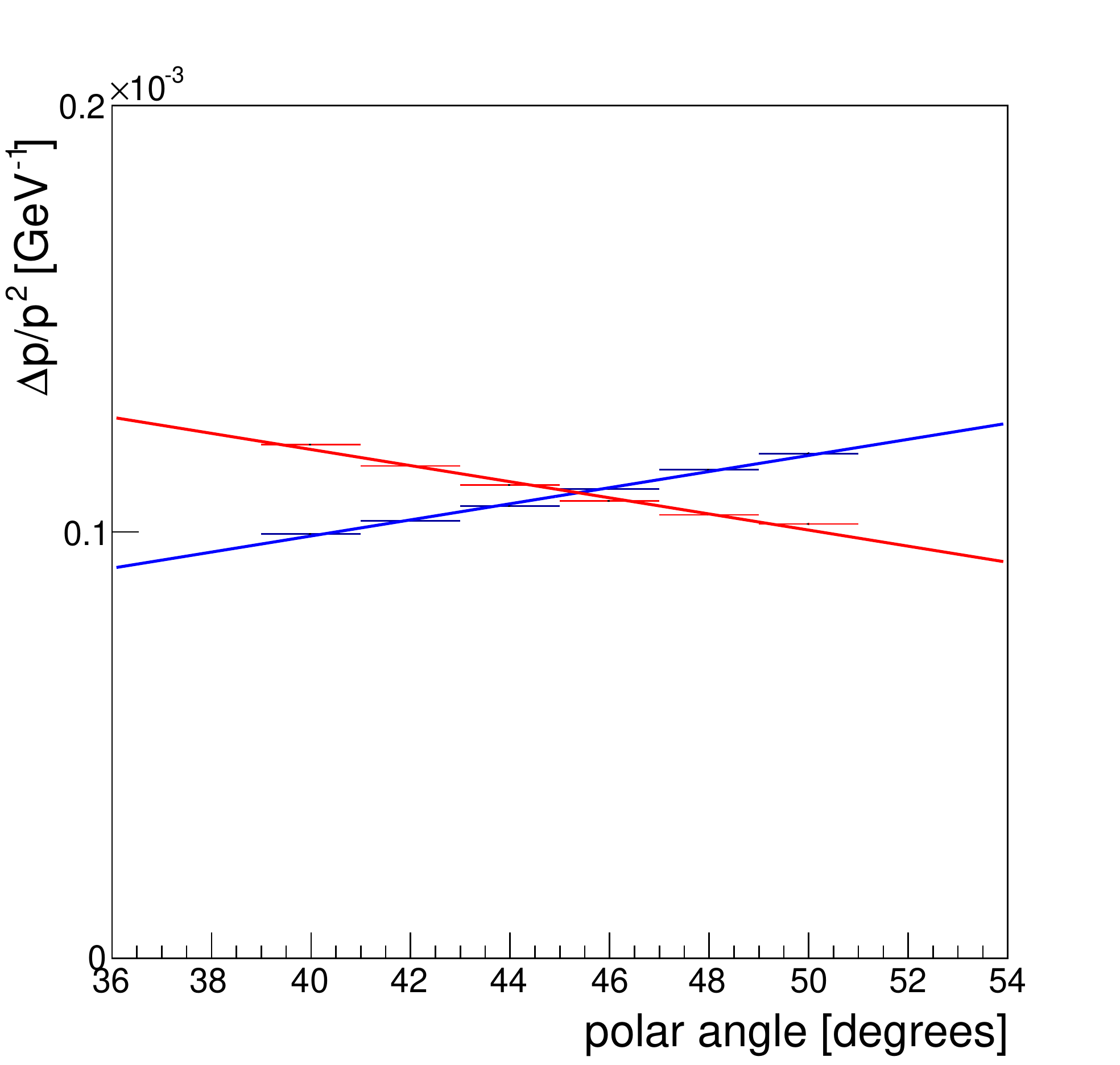}
\caption{The uncertainty on the transverse momentum (leftmost figure) and longitudinal momentum (rightmost figure) as a function of polar angle $\theta$. The results correspond to 100 GeV tracks emitted at a polar angle of close to 45$^\circ$ and measured in a toy detector, consisting of equally spaced cylindrical layers (blue lines) or disks (red lines). For $\theta = 45^\circ$ both setups yield identical results.}
\label{fig:barrel_vs_disks}
\end{center}
\end{figure}

Using the full 3-D information the trajectory of the particle at 
$\theta = $ 45$^\circ$ in our example is equally constrained by the 
barrel and end-cap setups. We find, using the CMS Kalman filter track 
fitter on a toy geometry, that this is
indeed the case. The track parameter uncertainties in 
Figure~\ref{fig:barrel_vs_disks} at $\theta = 45^\circ$ are
identical whether the track fit is fed with $r\phi$ and $z$ measurements
on horizontal ({\em barrel}) layers or $r \phi$ and $r$ measurements 
on vertical disks. 

The analytical approach with projections on the $r\phi$ and $rz$ planes has 
ceased to be relevant for the design of the experiment. There is no longer 
a need for dedicated algorithms to propagate or fit tracks in the forward 
detector.

\section{Environment: background}
\label{sec:backgrounds}

When two bunches of the ILC collide, beam particles radiate photons 
under the influence of the electromagnetic field of the opposite beam. 
These beamstrahlung photons convert into electron-positron pairs 
through several processes known as coherent and incoherent pair 
production. The interaction of the colliding bunches is modeled using a 
dedicated generator GuineaPig~\cite{guineapig1,guineapig2,guineapig3}.

The electrons and positrons produced in the interaction between the colliding 
bunches are generally emitted at small angle with respect to the beam axis
 and with small momenta. In the intense magnetic field in the tracking volume, 
the large majority of background particles curl up in spirals with a small 
radius. If the downstream structures are carefully designed
coherent pair production has a negligible contribution to the hit density
of the detector. Incoherent pair production, on the other hand, produces 
a small but significant fraction of the electrons and positrons 
with a transverse momentum exceeding 100 MeV. The number of produced pairs 
and their $p_T$ spectrum vary strongly
 with the final focus parameter and the center-of-mass energy, 
leading to an uncertainty of a factor 3--5 in the number of particles
reaching the innermost elements of the tracker and vertex detector. 

The $\gamma \gamma \rightarrow \mathrm{hadrons}$ process produces 
charged particles at a much lower rate than the pair production processes.
Due to the much harder momentum spectrum, however, 
charged particles can reach the outer layers of the detector, where 
this process is the dominant background source.

\begin{figure}[h!]
\begin{subfigure}{0.25\columnwidth}
\begin{tiny}
\begin{tabular}[b]{cc} \hline
\textbf{Detector} & \textbf{Hit density}  \\
\textbf{element} &  \textbf{($\mathrm{hits}/ \mathrm{mm}^2 / \mathrm{BX}$)}  \\ \hline
VXD1  & $ 3.2 \times 10^{-2} $ \\ 
VXD6  & $ 2.4 \times 10^{-4} $ \\ 
SIT2  & $ 4.0 \times 10^{-5} $  \\ 
FTD1  & $ 10^{-3}$ - \\
      & $ 10^{-5}$ \\            
FTD7  & $ 1.0 \times 10^{-5} $ \\ 
\end{tabular}
\end{tiny}
\label{tab:table1}
\end{subfigure}
\begin{subfigure}{0.75\columnwidth}\centering
\hfill \includegraphics[width=0.95\columnwidth]{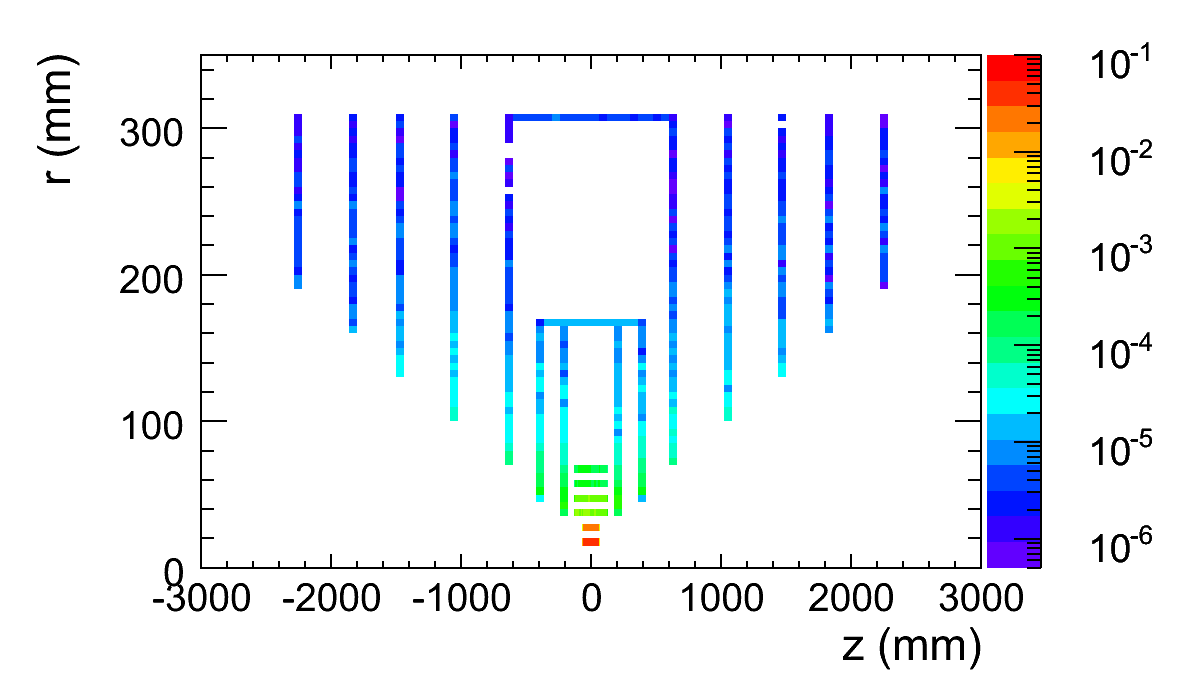}
\end{subfigure}
\caption{Background levels due to incoherent pair production in the inner tracking system of the ILD detector concept for 500 GeV operation of the ILC (in $\mathrm{hits/mm^2/BX}$). The table shows average densities over the detector surface, except for the innermost vertex detector disk (VXEC1) and the innermost forward tracking disk (FTD1), where the densities at the innermost and outermost 
radius are shown separately. The figure is reproduced from 
Reference~\cite{ildloi} and is based on the nominal set of final focus 
parameters.}
\label{fig:background_map}
\end{figure}

The direct hits due to electrons or positrons with relatively large transverse 
momenta are complemented by hits due to particles reflected off the material 
that surrounds the beam pipe further up- or downstream (small angle calorimeter 
systems). The result of a complete GEANT4~\cite{geant4} simulation of 100 
bunch crossings in the ILD detector at the ILC is shown 
in Figure~\ref{fig:background_map}.

The forward region of the tracking system receives a non-negligible amount 
of background hits. The background hit density in the inner rings of the 
first three disks is of the order of several times 
$10^{-4} \mathrm{hits}/\mathrm{mm}^2/$BX, 
comparable to that of the outermost vertex detector layers. 
Typically half of the hits on the forward tracking disks are due to 
particles reflected from upstream structures.

The occupancy due to background depends on the detector integration 
time with respect to the bunch structure of the collider. The 
envisaged bunch structures of ILC and CLIC are very different. For the ILC 
``cold technology'' based on superconducting cavities it is foreseen that 
every 200 ms there is a very brief ($\sim $ 1 ms) bunch train consisting of 
1312 bunches colliding at regular intervals of 500 ns. Existing 
detector concepts with a time resolution smaller than $\sim$ 500 ns 
can therefore distinguish individual bunch crossings at the ILC.

\begin{figure}[h!]
\begin{subfigure}{0.25\columnwidth}
\begin{tiny}
\begin{tabular}[b]{cc} \hline
\textbf{Detector} & \textbf{Hit density}  \\
\textbf{element} &  \textbf{($\mathrm{hits}/ \mathrm{mm}^2 / \mathrm{BX}$)}  \\ \hline
VXD1  & $7 \times 10^{-3}$ \\ 
VXD6  & $8 \times 10^{-4}$ \\ 
SIT2  & $7 \times 10^{-5}$ \\ 
VXEC1  & $7 \times 10^{-3} $  \\
      & - $2 \times 10^{-4}$ \\ 
FTD1  & $4 \times 10^{-3} $  \\
      & - $1 \times 10^{-4}$ \\ 
FTD5  & $9 \times 10^{-5}$ \\ 
      & - $3 \times 10^{-5}$ \\
\end{tabular}
\end{tiny}
\label{tab:table2}
\end{subfigure}
\begin{subfigure}{0.75\columnwidth}\centering
\includegraphics[width=0.95\columnwidth]{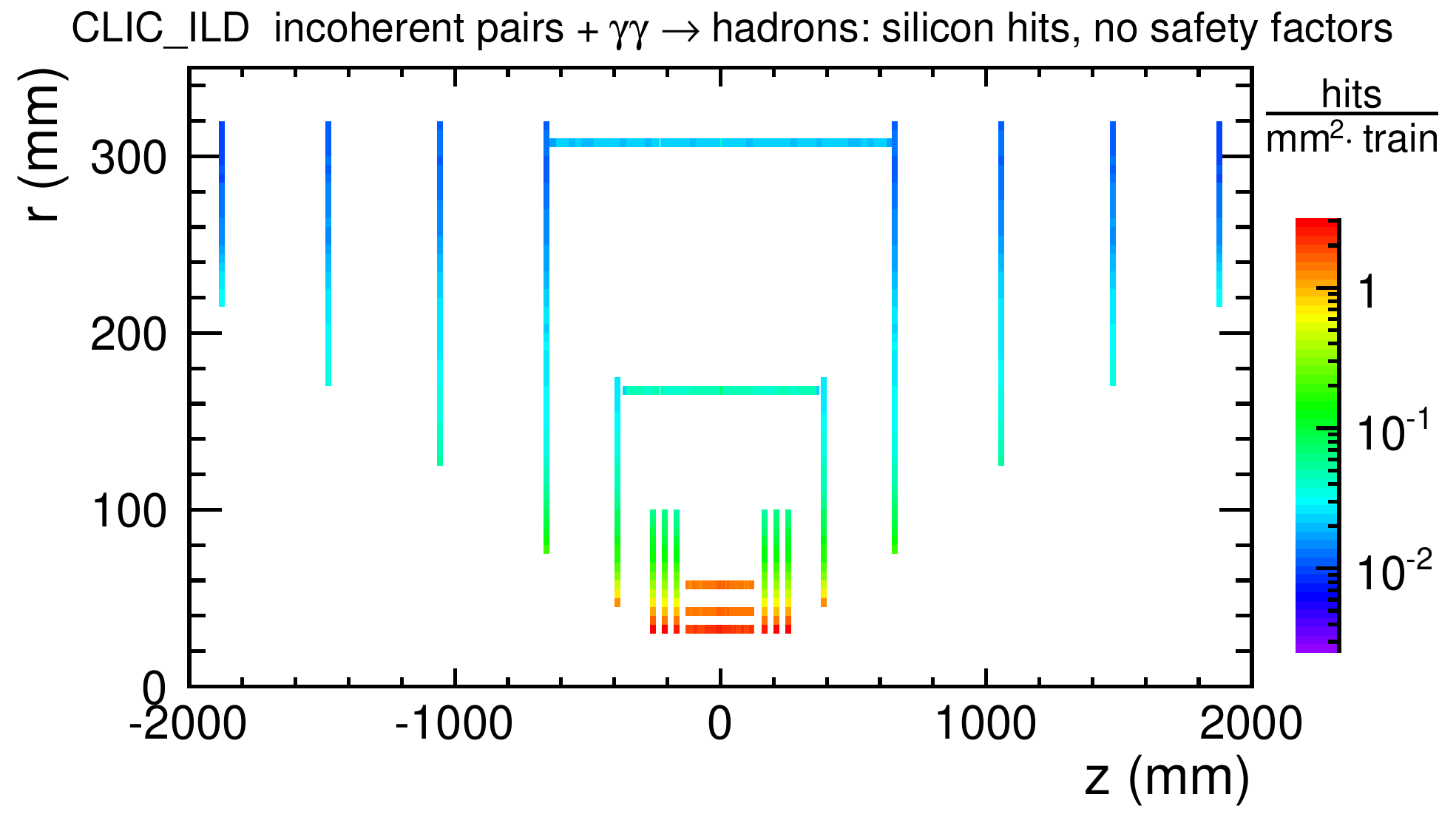}
\end{subfigure}
\caption{The background hit density (in $\mathrm{hits/mm^2/}$bunch train) 
in the innermost 
tracking system of the ILD detector in the CLIC environment at $\sqrt{s}=$ 
3 TeV. The contributions of incoherent pair production and 
$ \gamma \gamma \rightarrow \mathrm{hadrons}$ are added. The table shows
average densities over the detector surface, except for VXEC1, FTD1 and FTD5, 
where the densities at the innermost and outermost radius
are shown separately.}
\label{fig:background_map_clic}
\end{figure}

The hit density in the CLIC-ILD tracker layers due to incoherent pair production  
and $\gamma \gamma \rightarrow \mathrm{hadrons}$, based on the study of 
Reference~\cite{lcd-note-2011-021}, is shown in 
Figure~\ref{fig:background_map_clic}. The layout of the innermost layers 
is modified with respect to the background map for ILC in 
Figure~\ref{fig:background_map}. In particular, the innermost
layer was moved to $r=$ 3.1 cm, instead of the 1.5 cm at the 
ILC~\cite{lcd-note-2011-031}. The CLIC bunch structure is is different
as well; every 20 ms a short train of 312 bunches at 0.5 ns intervals 
collides. Sub-bunch train time-stamping at CLIC requires 
new read-out concepts to be developed. The results are therefore 
presented as the hit density per train of 312 bunch crossings.  
And, finally, the contribution of 
\mbox{$ \gamma \gamma \rightarrow \mathrm{hadrons}$} is included for CLIC. 
The contribution of incoherent 
pair production is dominant in all layers of the inner tracking
system except for the barrel tracker (SIT) and for outer radii of
the forward tracking disks, where the 
contribution of $ \gamma \gamma \rightarrow \mathrm{hadrons}$ reaches 
a similar level~\cite{lcd-note-2011-021}.

The hit density in the forward tracker presents a constraint and a 
challenge to the design. To cope with the background levels a combination 
of fast read-out and high detector granularity is required.

\section{Momentum resolution}
\label{sec:momentum}

The recoil-mass method to determine the couplings of the Higgs boson 
is a central piece of the LC physics programme. It requires excellent momentum 
reconstruction for muons with $E \sim (\sqrt{s} - m_H) / 2$~\cite{Li:2009fc}. 
In a high-energy collider, a similar requirement is derived from the analysis 
of the $H \rightarrow  \mu^+ \mu^-$ decay~\cite{Grefe:2012eh,Grefe:2012gj}. 
Therefore, all LC detector concepts aim for excellent momentum resolution 
for charged tracks, usually specified as:
 \begin{equation}
 \Delta  \Bigg( \frac{1}{p} \Bigg) \/ \mathrm{[GeV]}^{-1} = 2 \times 10^{-5}  \oplus \frac{10^{-3}}{p \/ [\mathrm{GeV}]} 
\label{eq:momentumspec}
\end{equation}
The first term applies to the asymptotic resolution for high momentum
tracks, where multiple Coulomb scattering in the detector material is 
negligible. The second term constrains the material budget of the 
tracker. At this level the tracker resolution has negligible impact
on the energy resolution of the particle flow algorithm: 
$\Delta p_T/p_T \ll \Delta E_j/E_j$ for all relevant momenta.

Compared to previous experiments, the momentum resolution requirement 
of the linear collider experiments implies a very significant improvement. 
The LEP experiments reached 
$5 \times 10^{-4} $~GeV$^{-1}$~\cite{aleph_detector,delphi_detector}. 
The LHC general-purpose experiments quote a relative transverse momentum 
resolution $ \sigma (p_T) / p_T $ for central 100 GeV particles in the range 
from 1.5 \% (CMS~\cite{cmsphysics}) to 4 \% (ATLAS~\cite{cscnote}), 
corresponding to $ \Delta (1/p_T ) = 1-4 \times 10^{-4} $~GeV$^{-1}$.

In the following we discuss the dependence of the transverse momentum 
resolution on the polar angle. An important caveat must be kept in mind. 
The ability of the experiment to reconstruct narrow resonances decaying to 
pairs of charged particles depends on the precision of the total momentum
measurement (rather than on that on the transverse momentum). Indeed, the total
momentum is the relevant quantity for most physics analyses. It is 
therefore more reasonable to aim for a uniform resolution for the $(1/p)$
measurement, than to require a constant $\Delta (1/p_T)$ over
the whole detector. As the 
polar angle at the production vertex is measured with excellent 
resolution the uncertainty on the total momentum is generally dominated 
by that on the transverse momentum: $\Delta p_T/p_T \sim \Delta p/p$. 
While in the central detector $\Delta (1/p) = \Delta p/p^2$
and $\Delta (1/p_T) ) = \Delta p_T/p^2_T$ are identical, in the forward 
region both quantities differ by a factor $(\sin{\theta})^{-1}$ (in the limit
of perfect $\theta$ measurement). This factor can be
quite important, i.e. for tracks emitted at 20$^\circ$ the two quantities
differ by nearly a factor three. 

\subsection{Performance \& detector design}

\begin{figure}
\centering
\includegraphics[width=0.49\columnwidth]{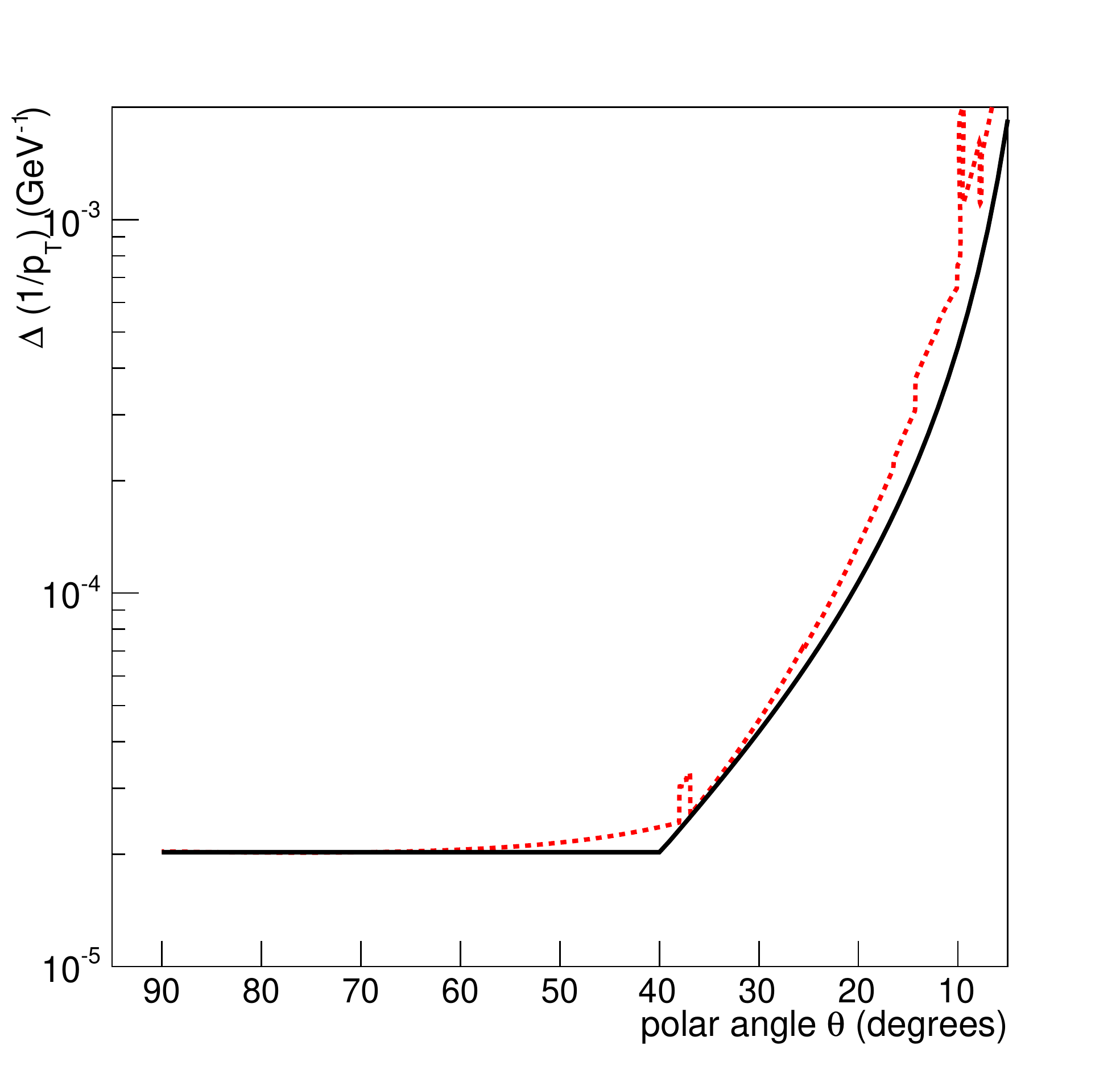}
\includegraphics[width=0.49\columnwidth]{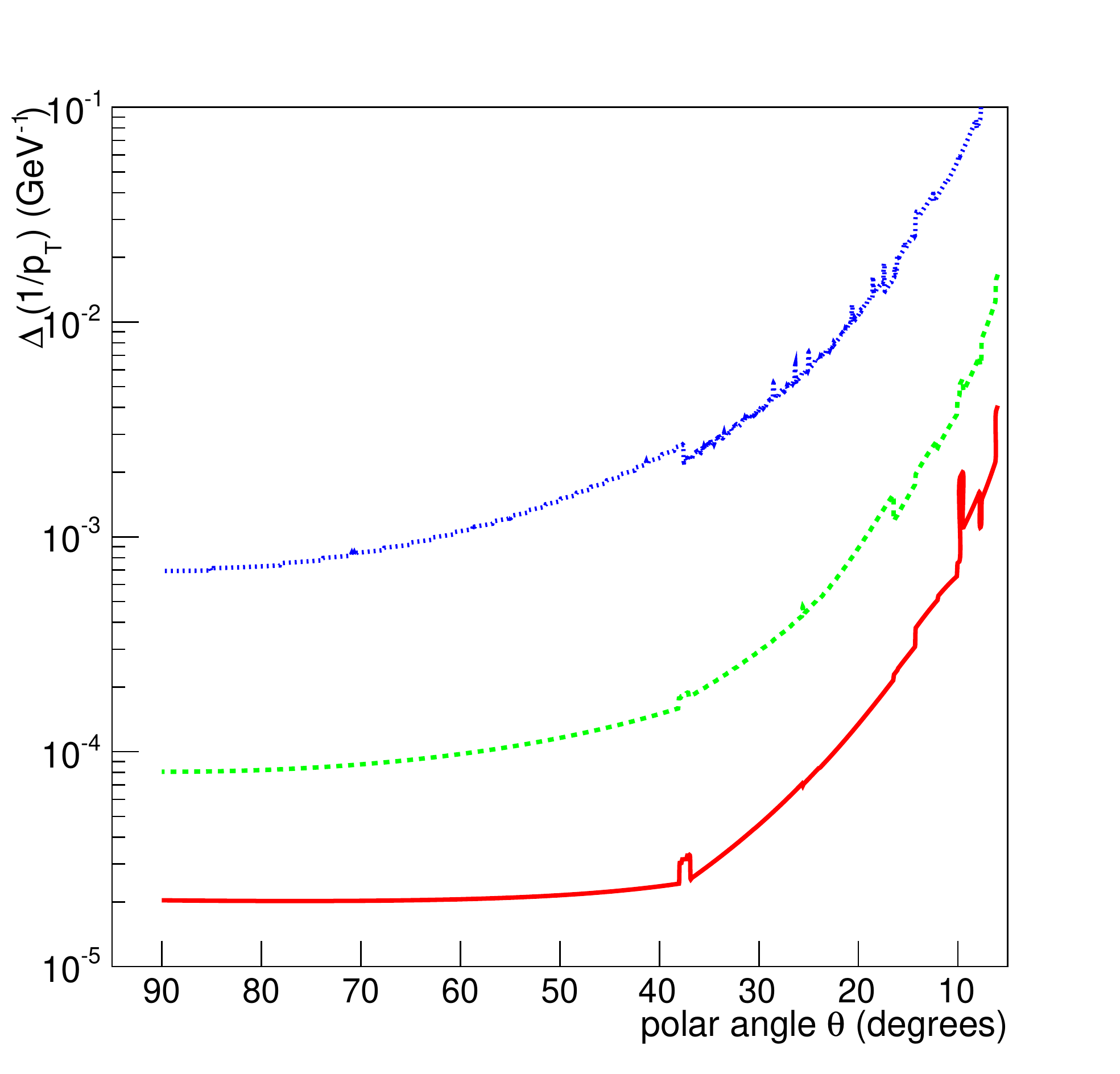}
\caption{The transverse momentum resolution versus polar angle $ \theta $ for single muons in the ILD detector, obtained with the LiC detector toy fast simulation~\cite{lictoytheprogram,lictoyapplicationild}. In the leftmost panel the result for 100 GeV muons (dashed line) is compared to the prediction of the Gluckstern formula (continuous line). The spatial resolution of the latter is chosen such that the performance of the central detector is in good agreement. The three graphs in the rightmost panel correspond to momenta of 1 (blue dotted line), 10 (green dashed line) and 100 GeV (red continuous line).}
\label{fig:trackingperf_pt}
\end{figure}

To expose the dependence 
on the parameters of the tracker design we consider an analytical approximation
before turning to the simulation.
The asymptotic momentum resolution for high-energy charged particles 
of a tracker consisting of N equally spaced
layers\footnote{The Gluckstern formula is valid for $N>$10. But, even 
for only five space points involved in the track fit of the most forward 
particles, the exact numerical constant approaches $\frac{720}{N+4}$ to
within 7~\%.} with spatial resolution $ \sigma_{r \phi}$  in a magnetic 
field $B$ (in Tesla) and with lever arm $L$ perpendicular to the magnetic field
 (in meters) is given 
by~\cite{Gluckstern:1963ng}:
\begin{equation}
\frac{\sigma (p_T)}{p_T} = \sqrt{\frac{720}{N+4}} \sigma_{r \phi} \frac{p_T}{0.3 B L^2},
\label{eq:gluckstern}
\end{equation}

Assuming uniform quality of the instrumentation (i.e. all cylindrical layers and
disk provide measurement with the same resolution $\sigma_{r \phi}$ 
and adopting the aspect ratio of the ILD experiment, the 
asymptotic transverse momentum resolution in the absence of multiple 
scattering for 100 GeV tracks has the polar angle
dependence shown in the leftmost panel of Figure~\ref{fig:trackingperf_pt}. 
We find a considerable 
degradation of the performance, over more than an order magnitude, 
towards smaller polar angle. This is entirely due to the reduction 
of the projected length (on the transverse plane) of the trajectory $L$.

Superposed in the same Figure is the LiCToy result for the approximate ILD 
geometry. The curves follow each other rather closely for polar angles
within the TPC acceptance (down to approximately 25$^\circ$). This is 
indication that the design provides rather uniform quality. In ILD
this is achieved due to the partial cancellation of two effects; the decreasing 
number of space points in the TPC leads to a degradation at small polar angle
that is more pronounced than the Gluckstern prediction in  
formula~\ref{eq:gluckstern}. 
The spatial resolution of the TPC depends, however, on the drift distance 
through the gas volume. The parameterization in the simulation has~\cite{ildloi}:
\begin{equation}
\sigma (r \phi) [\mu \mathrm{m}] = 50 \mu\mathrm{m}  \oplus 900 \mu\mathrm{m} \sin{\phi}  \oplus 28 \mu\mathrm{m} \sin{\theta} \times \sqrt{\Delta z} [\mathrm{cm}]
\end{equation}
where $\Delta z$ is the drift distance in centimeters.
The resolution of space points close to the endplate is 
much improved. Therefore the forward performance recovers partially.

The momentum determination of the most shallow particles rests primarily 
on the measurements in the Forward Tracking Disks. For polar angles below 
16$^\circ$ the number of measurements is reduced to 4-6. The
simulation envisages a spatial resolution of 7 $\mu \mathrm{m}$, which is
clearly insufficient to stay on the Gluckstern curve. 

The discussion so far has focused on relatively high momentum tracks. 
The performance of a realistic tracker design for particles with a momentum 
below several tens of GeV is dominated by multiple scattering and thus
depends crucially on a tight control of the detector material.
LiCToy results for the transverse momentum resolution of particles with 
momenta of 100 GeV, 10 GeV and 1 GeV in the ILD detector are compared in 
Figure~\ref{fig:trackingperf_pt}. In agreement with the functional form
of equation~\ref{eq:momentumspec} the resolution for 10 GeV particles 
is degraded by a factor 4 and that for 1 GeV particles by a factor 40.
In the ultra-transparent tracker design of ILD the material term in
 equation~\ref{eq:momentumspec} becomes comparable to 
the first term for $p \sim $ 50 GeV, the energy of muons from $Z$-decay in 
$ZH$ production at $\sqrt{s} = $ 250 GeV. 

\subsection{Discussion}

The parameters that govern the momentum resolution for forward tracks
are highly constrained by the overall detector layout
(and cost). A factor two improvement of the momentum resolution 
requires to double the magnetic field or the length along $z$ 
of the tracking volume. Neither of these are viable
in the ILD and SiD design, where the total cost of the experiment
receives large contributions from the coil (whose cost is approximately
proportional to the total stored energy) and the calorimeter system
(whose cost scales roughly with the length).  
Additional measurement layers do provide a means to improve the momentum
resolution for very high momentum tracks, but the extra material
severely damages the ability of the experiment to precisely
reconstruct the abundant low momentum tracks in the forward region.
The parameter that remains is the precision of the space points. 
To achieve the best possible momentum resolution, detector R\&D must
pursue devices that provide better $r \phi$ resolution 
without increasing the material budget.

\section{Vertex reconstruction}
\label{sec:vertex}

 The identification of heavy flavour jets through the displaced vertex of 
long-lived beauty and charm hadrons has proven to be a crucial technique in 
previous experiments. Even for 500 GeV to 1 TeV operation charged
particles carrying the lifetime information in the jets are often quite soft. 
Therefore, flavour tagging depends crucially on the precise reconstruction of 
low momentum ($ < $ 10 GeV) tracks, where multiple scattering dominates the 
performance. 

\subsection{Performance \& detector design} 

The ILC detector concepts have drawn up severe requirements on the vertexing 
performance of their tracking systems. The goal is usually expressed in 
terms of the following parameterization of the transverse impact parameter 
resolution in a cylindrical vertex detector with uniform spatial
resolution and material:
\begin{equation} 
  \Delta d_0  [\mu\mathrm{m}] = a \/ [\mu \mathrm{m}] \oplus \frac{b \/ [\mu \mathrm{m}] }{p \/ [\mathrm{GeV}] \sin^{3/2}{\theta}}  
\label{eq:impactparameter}
\end{equation}
The LC concepts have formulated a goal of $a=5$, $b=10$. The 
value of the material term $b$ is increased to 15 in 
the CLIC requirement. To achieve excellent flavour tagging a similar 
resolution on the longitudinal impact parameter is required,
 as well as excellent two-track resolution.

The transverse impact parameter resolution requirement in 
equation~\ref{eq:impactparameter} represents a 
considerable improvement over vertex detectors built at 
collider experiments to date; the constant term is better by a factor 2--4 than 
what was achieved at previous $e^+ e^-$ colliders and at the LHC. Achieving
the requirement for the second (material) term is even more challenging; 
it has to decrease by a factor 6--10 with respect to most 
previous experiments. Indeed, the requirement in 
Formula~\ref{eq:impactparameter} (together with the assumed inner radius of 
15 mm) implies that the vertex detector must be built with a strict 
material budget of order 0.1\% of a radiation length per layer, a factor 
three better than the best material term achieved so far (by the SLD 
vertex detector) with:
\begin{equation}
 \Delta d_0  [\mu\mathrm{m}] = 9 [\mu \mathrm{m}] \oplus \frac{33 [\mu \mathrm{m}]}{p [\mathrm{GeV}] \sin^{3/2}{\theta}}. 
\end{equation}

The ${\sin^{-3/2}{\theta}}$ form for the polar angle dependence of the 
the impact parameter resolution (the second term in 
equation~\ref{eq:impactparameter}) can be understood as follows.
The distance of the innermost barrel layer to the interaction point increases 
as $\sin{\theta}$. Multiple scattering is proportional to the square root
of the material thickness in radiation lengths~\cite{Highland:1975pq}. 
For an ideal detector of constant thickness this is proportional to 
the path length through the detector, that grows as $\sin{\theta}$. 
Following the same argument we can rewrite equation~\ref{eq:impactparameter} 
for the case of polar angles where the first measurement is on an end-cap disk:
\begin{equation} 
  \Delta d_0 [\mu\mathrm{m}] = a \/ [\mu \mathrm{m}] \oplus \frac{b [\mu \mathrm{m}] \times \frac{L}{R}   }{p \/ [\mathrm{GeV}] \cos^{3/2}{\theta}}.  
\label{eq:impactparameterfwd}
\end{equation}
Due to the different orientation of the disks the $\sin^{-3/2}{\theta}$ 
dependence is replaced by a $\cos^{-3/2}{\theta}$ form. The numerical
constant $b$ of equation~\ref{eq:impactparameter} is multiplied by
the ratio $\frac{L}{R}$ of the distance $L$ (along $z$) of the disk 
to the interaction point and the inner radius $R_i$ of the barrel 
detector. Note that equation~\ref{eq:impactparameterfwd} ignores the
contribution of the beam pipe. 

\subsection{Simulation} 

\begin{figure}[h]
\centering
\includegraphics[width=\columnwidth]{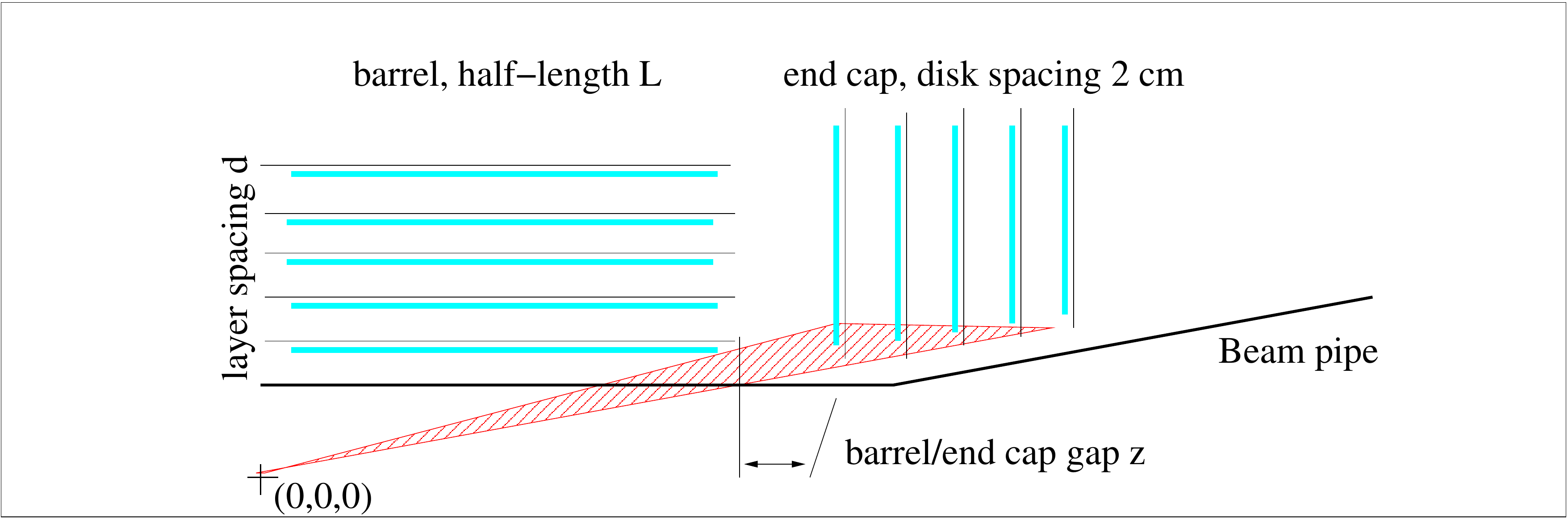}
\caption{The generic vertex detector layout. The choice of parameter is described in the text.}
\label{fig:vertex_toy}
\end{figure}

We evaluate the performance 
of the simplified layout shown schematically in figure~\ref{fig:vertex_toy}. 
The most important parameter of the detector design is the inner radius of 
the innermost barrel layer and the disks. In particular the material term $b$ 
is proportional to the inner radius~\cite{lcd-note-2011-031}. In the ILC 
experiments the minimal distance is limited to approximately $R \geq $ 
15 mm by the pair background. The CLIC vertex design envisages an inner radius
of approximately 3.1 cm. To avoid the envelope 
of the intense core of pair production the beam pipe of the LC concepts has 
a conical shape beyond a certain $|z|$-value, such 
that in practice the minimum inner radius depends on the z-extension of the 
barrel or z-position of the disks. 

The spatial resolution of the $r \phi$ and z (r) measurements of each layer 
(disk) is set to 3 $\mu \mathrm{m}$, the material to 0.12 \% of a radiation 
length per layer. The gap between barrel and end cap structures must be 
minimized for optimal performance, within the boundary conditions due to 
mechanics and services. We consider $z=$ 1 cm. Finally, the spacing between
layers has relatively little impact on the performance and is fixed 
to $d=$ 0.8 cm and $d$ = 2 cm, in barrel and endcap, respectively.

\begin{figure}
\centering
\includegraphics[width=0.49\columnwidth]{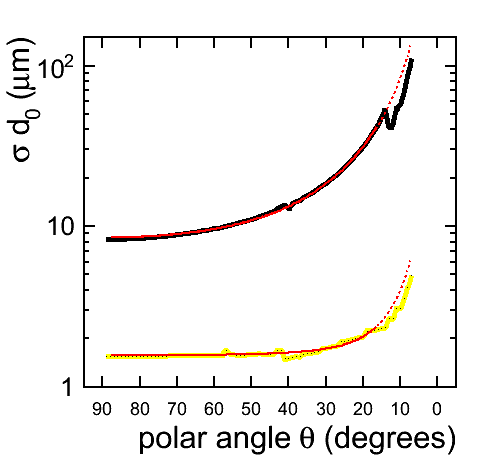}
\includegraphics[width=0.49\columnwidth]{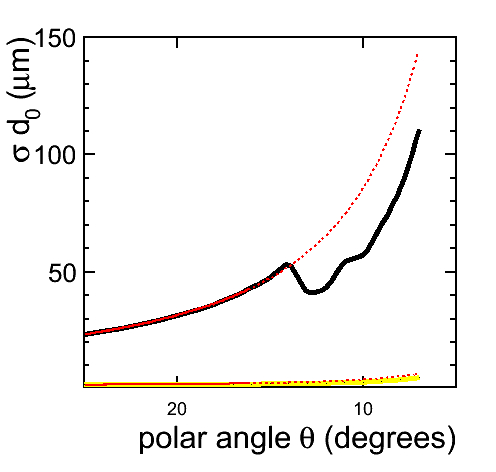}
\caption{The transverse impact parameter resolution versus polar angle 
$ \theta $ for single tracks in the toy vertex detector specified in the text. 
The two curves in the leftmost panel correspond to particles with a 
momentum of 1 GeV (black) and 100 GeV (yellow or grey). The functional form of 
equation~\ref{eq:impactparameter} is fit to the result in the interval 
$ 90^\circ < \theta < 17^\circ $ (red or dashed curves). The rightmost panel
zooms in on the forward region.}
\label{fig:trackingperf_d0}
\end{figure}

The transverse impact parameter resolution of the ILC toy detector with 
$R_i = $ 1.5 cm is presented in Figure~\ref{fig:trackingperf_d0}.  
The two curves correspond to particles with a momentum of 1 GeV (black) and 
100 GeV (yellow/grey) and thus give an estimate of the $b$ and $a$ parameters,
respectively. The functional form of equation~\ref{eq:impactparameter} is fit 
to the result in the interval $90 < \theta < 17^\circ$ (red/dashed curves).
The parameterization yields an adequate description of the observed curve 
in the central region of the detector (this holds also for full simulation 
results). With these aggressive design assumptions the LC requirements 
are met in the central region. The parameter $a$ in the constant term 
is less than 2, well below the required value of 5. 
A $b$ parameter in the material term of 8 is achieved, 
again better than the required value of 10.

The functional form of equation~\ref{eq:impactparameter} implies a
steady degradation of the vertexing performance towards the forward region,
that is inherent in the cylindrical layout of the barrel vertex detector. 
The degradation is certainly significant for 100 GeV particles, but
the $a$ term remains below 5 for the complete polar
angle range considered. The material term $b$ increases much more strongly. 
The combination of the larger extrapolation distance with the increase 
in the amount of material traversed by particles at shallow angle, has 
a strong impact on the performance. The uncertainty on $d_0$ for low 
momentum particles doubles from 8 to 16 $\mu m$ before reaching 30$^\circ$ 
and doubles again between 30$^\circ$ and 20$^\circ$. 

The parameterization of equation~\ref{eq:impactparameterfwd} for particles
that leave a first hit on an end-cap disk is found to be in qualitative 
agreement with the simulation. The zoom image of the forward region 
in the rightmost panel of Figure~\ref{fig:trackingperf_d0} indeed shows 
how the resolution improves sharply beyond the barrel-endcap transition
at approximately 15$^\circ$. This is primarily due to the reduced 
path length through the detector material. As the polar angle is reduced 
further the resolution improves until the coverage of the first disk is 
insufficient and a disk at greater $\mathrm{z}$-distance from the
interaction point takes over.

\subsection{Discussion} 

One can see from the results of this simple model that a detector with a 
short barrel cuts off the $ \sin^{-3/2}{\theta}$ growth of the impact 
parameter resolution early and
provides better performance at any polar angle. The rationale
for the {\em long barrel} geometry adopted by ILD only becomes clear when more 
realism is added to the material budget. Even a minor contribution to
the material budget in the conical region marked with a (red) fill pattern
in Figure~\ref{fig:vertex_toy} leads to a severe degradation of the
impact parameter of (low momentum) tracks emitted in that direction.
It is noted here, however, that the long barrel layout cannot provide
full polar angle coverage. The forward disks in this design are too
distant from the interaction point to provide good performance. 
The vertexing performance for very shallow tracks
is thus sacrificed.

The desire to keep the material in the ``no-go'' cone to an absolute minimum
leads to a number of engineering challenges. 
Clearly, the beam pipe, that is traversed at an unfavourable angle,
must be kept as thin as possible. A potentially even bigger threat are the
services and support structures of the barrel vertex detector. To maintain
good vertexing performance for particles that leave their first hit on
the end-cap it is vital to route the services of the barrel over the end-cap.
Also the supports and end-of-ladder area of the barrel vertex detector 
must avoid the ``no-go'' cone as much as possible.

\section{Pattern recognition} 
\label{sec:pattern}

The particle flow paradigm requires unambiguous association of all
charged particle trajectories to energy deposits in the calorimeter.
Confusion due to reconstruction inefficiency, badly reconstructed tracks 
or accidental combinations of hits yielding {\em fake} particle candidates
must be kept to a minimum. The tracker at a linear collider
experiment must therefore be capable of excellent pattern recognition.

A number of reasons render pattern recognition in the forward region of 
the detector more challenging than in the central detector:
\begin{itemize}
\item As discussed in Section~\ref{sec:backgrounds} the hit density due to
beam-induced background drops rapidly with increasing radial distance
from the interaction point. The forward tracking devices closest to the 
beam pipe must therefore cope with severe background levels. 
This is further amplified by other background sources with a markedly forward 
profile like $\gamma \gamma \rightarrow \mathrm{hadrons}$ production. 
\item The strong magnetic field leads to an abundance of low momentum 
particles that leave the tracking volume curling through the 
forward tracking region (so called {\em loopers}). These represent
a challenge to pattern recognition.
\item The LC detector design have elongated tracking volumes, with an 
aspect ratio of approximately 75\%. The larger distance between layers 
leads to a larger uncertainty on the extrapolated position on the next layer.
\end{itemize}

In this section we relate pattern recognition requirements to the detector
design. We do not attempt to derive an analytic expression, as we did 
for the momentum resolution. Pattern recognition is, however, 
important in the formulation of the detector specifications. We present 
quantitative arguments in Section~\ref{sec:occ}, based on the hit density 
of signal and background events, that motivate the choice for 
high-granularity devices in the innermost layers. We discuss
the formation of ghost hits in $\mu$-strip detectors in 
Section~\ref{sec:ghosts} and put forward a detector design that
minimizes their impact. We also discuss, in Section~\ref{sec:strips}, the 
precision that should be required of the $r$-measurement of the tracking 
layers.


\subsection{Detector occupancy}
\label{sec:occ}

The foremost marker to optimize the detector granularity is the detector 
occupancy. At the LC it is safe to assume a single signal or physics event 
contributes to the detector occupancy. To this signal 
occupancy we must add the contribution of a certain number of beam-induced 
background events. The exact number of overlaid bunch crossings depends
on the integration time and time stamping cababilities of the detector
and may vary from one sub-system to the next.
The total occupancy may be given as the sum of the occupancy from the physics 
event $O_P$ and the background occupancy $O_B$. We
determine a representative $O_P$ in $e^+ e^- \rightarrow t \bar{t}$ events. The
background density $O_B$ is based on Figure~\ref{fig:background_map}. 
For the innermost Forward Tracking Disk at the ILC one obtains the 
following average hit density: 
\begin{equation}
O_p + O_B = 1 \times 10^{-4} \frac{\mathrm{hits}}{\mathrm{mm}^2} + 1.6 \times 10^{-4} \frac{\mathrm{hits}}{\mathrm{mm}^2 \mathrm{BX}} (\mathrm{average}).
\label{eq:avgdensity}
\end{equation}
This number reflects the number of charged particles traversing the 
corresponding detector element per unit
area and time.  
It must be multiplied by the (average) number of channels that 
fires for each hit. As this is a technology-dependent quantity we
do not include this factor here. Note that a threshold is applied for 
the energy deposition in the detectors that is also (mildly) technology
dependent~\cite{Dannheim:2012mf}.

The hit density is subject to large event-to-event fluctuations.
In the innermost detector layers jets can provoke a local
occupancy that is two order of magnitudes larger than the average occupancy.
The average occupancy moreover depends 
strongly on the position in the detector.
The variation is particularly pronounced in the forward tracking disks,
where the innermost ring has to deal with 10 times more background than the
outer regions of the same disk. The peak hit density from
signal and background in the innermost Forward Tracking Disk at the ILC 
then becomes:
\begin{equation}
O_p + O_B = 1 \times 10^{-2} \frac{\mathrm{hits}}{\mathrm{mm}^2} + 1.6 \times 10^{-3} \frac{\mathrm{hits}}{\mathrm{mm}^2 \mathrm{BX}} (\mathrm{peak values}).
\label{eq:peakdensity}
\end{equation}
Locally, in the core of jets, the contribution of physics event can exceed the
background density from a single bunch crossing by an order of magnitude.

The relative contribution of the background hits depends on the technology 
chosen to equip the detector. At the ILC bunches cross every 500 ns. 
With fast electronics a single bunch crossing can be 
integrated and the high local density of signal events forms the 
tightest requirement. Micro-strip detectors are fast enough that 
single bunch crossings at the ILC can be identified. For 10 cm long, 
50 $\mu\mathrm{m}$ wide strips a peak occupancy of 6\% per BX is obtained, which
is well over the maximum (more on this in the discussion of ghost hits). 
Therefore, detectors with finer granularity
are required. Several candidate technologies~\cite{depfet,cmos} can
produce devices with a cell area of 25 $\times$ 25 $ \mu \mathrm{m}^2$ and
a read-out time of 50 $\mu \mathrm{s}$. The peak occupancy due to signal 
in the core of jets is less than $10^{-5}$, allowing for robust pattern 
recognition. When the background hits from approximately 100 bunch 
crossings are added, the total occupancy becomes $1 \times 10^{-4}$. 
The total occupancy remains at a comfortable level,
as the very small pixel area (reduced by nearly four orders of magnitude) 
compensates for the increased integration time (100 bunch 
crossings as opposed to a single bunch crossing). The LC detector 
concepts have therefore opted for pixel detectors in the 
innermost Forward Tracking Disks.

An alternative solution is offered by the fine pixel CCD 
detectors~\cite{fpccd} that integrate the 1312 bunch crossings of the 
bunch train. The occupancy is kept at an acceptable level by reducing
the pixel size to below 10 $\times$ 10 $\mu \mathrm{m}^2$.

At CLIC the bunch crossings are spaced by only 0.5 ns. 
It is assumed that the tracking and vertex detector integrate hits over 
the train duration of 156 ns. Time stamping with a precision of 10 ns
is sufficient to reduce the expected background occupancy to an acceptable
level~\cite{cliccdr2}. Low-mass and low-power hybrid pixel 
detectors with a pitch of approximately 
25 $\times $ 25 $\mu\mathrm{m}^2$~\cite{clicpix} and a read-out architecture
based on TimePix~\cite{timepix} are expected to achieve this requirement. 
Time stamping at the level of a single bunch crossing may be possible
with specially designed, ultra-fast detectors.

For very shallow tracks we cannot rely on the quiet layers at
large radial distance from the interaction point or the large number 
of measurements in the TPC.
The presence of closely spaced detectors, where track
candidates can be extrapolated to the following layer with an error
of less than 10 $\mu \mathrm{m}$, and low occupancy, so that few
new ambiguities arise, is crucial to sort out the ambiguities that
inevitably arise in track finding. To increase the robustness of the 
forward system at CLIC the innermost forward tracking system is 
therefore formed by six closely spaced and highly granular 
disks~\cite{lcd-note-2011-031}.



 
\subsection{Ghost hits}
\label{sec:ghosts}

 Micro-strip detectors remain the most 
economical option in terms of channel count and power density for 
large detector areas. In both cases the specification of the $r \phi$ resolution
is tightly constrained from the requirements on the momentum and transverse
impact parameter resolution. 

For micro-strip detectors in the forward region this argument naturally 
leads to the choice of radially oriented strips. To provide a precise 
measurement of the 
$r\phi$-coordinate that measures the track curvature the strips must be 
aligned with $z$ in the cylindrical layers and radially along $r$ in the 
disks. The constraint on the second coordinate ($z$ and $r$, respectively) 
detector is very poor, as we can only infer that the particle traversed a 
given strip (typical sensors produced from 8 inch wafers are up to 10 
cm long, often several sensors are ganged to produce even longer strips). 

\begin{figure}
\centering
\includegraphics[width=\columnwidth]{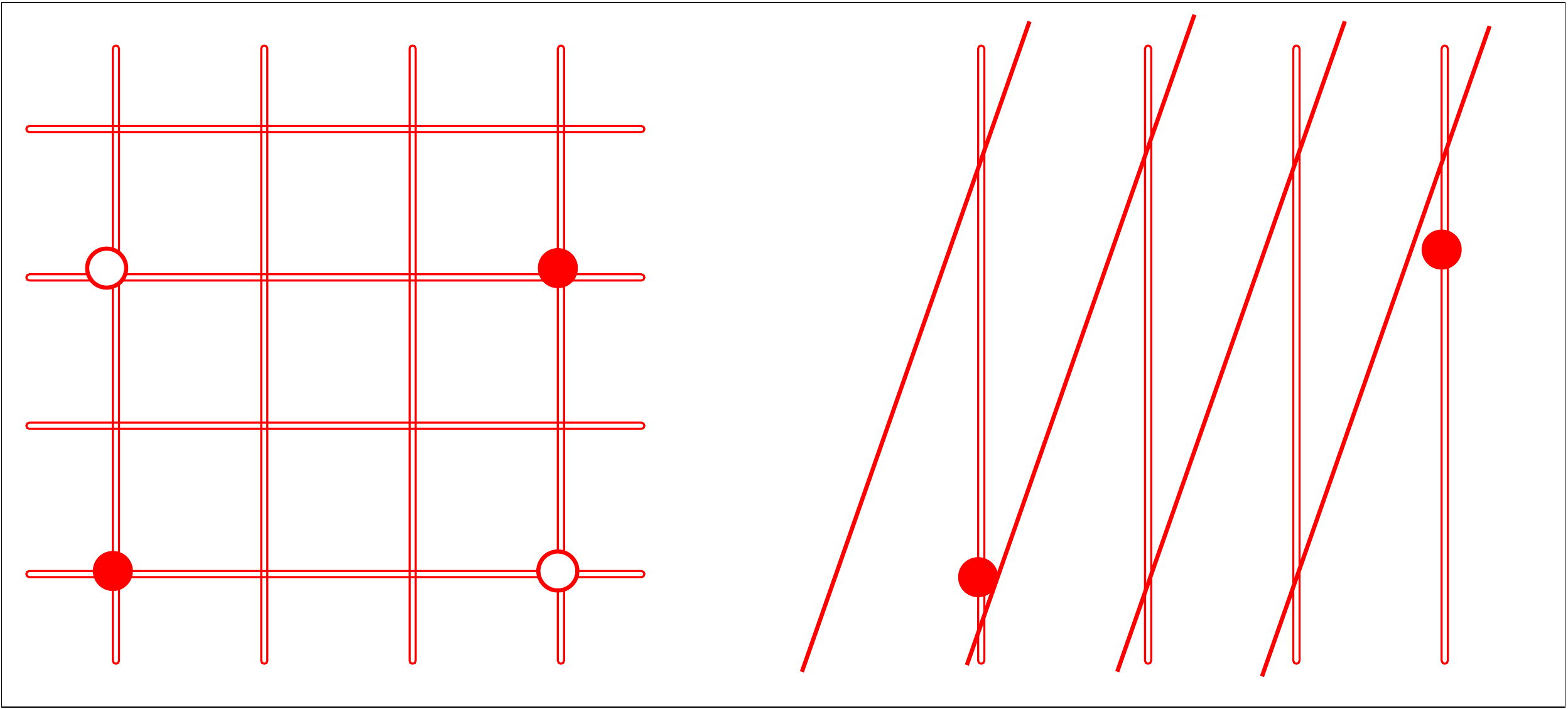}
\caption{Schematic representation of the formation of {\em ghost} hits in micro-strip detectors with 90 degree stereo angle (leftmost image). The incidence positions of two particles are indicated as filled red dots. Two additional valid combinations of the 1D information of both detectors are indicated as open circles. The rightmost image demonstrates how detectors at a 10 degree stereo angle produce no ghost hits in the same situation. }
\label{fig:ghosts_scheme}
\end{figure}

A more precise determination of the second coordinate is obtained if two 
measurements are combined. If the second sensor is rotated by 90$^\circ$, 
the second coordinate is measured with the same resolution as the primary
$r\phi$ measurement. This configuration has, however, rather poor 
performance in a dense environment. Whenever more than one particle traverses
a sensor, several {\em ghost} combinations appear that cannot be distinguished
(using local information) from the three real hits. An example is shown
in Figure~\ref{fig:ghosts_scheme}. The total number of 
valid hit combinations scales with $N^2$, where $N$ is the number of particles.

A solution to the problem of the ghost hits, adopted for example by the LHC 
experiments, is to use a small stereo angle of 40--100 milliradians. 
The combination of the two stereo measurements yields a determination of
the second coordinate, with a precision that depends on the stereo angle
(better resolution is achieved at larger angle). A more extensive discussion
of the resolution is left for Section~\ref{sec:strips}.
The rightmost panel of Figure~\ref{fig:ghosts_scheme} shows that no
ambiguities arise in the same situation as before if the strips are rotated
by a 10 degree stereo angle. As each micro-strip overlaps with only
a limited number of stereo strips, the probability to find a second hit in
an overlapping stereo strip is greatly reduced.

\begin{figure}
\centering
\includegraphics[width=\columnwidth]{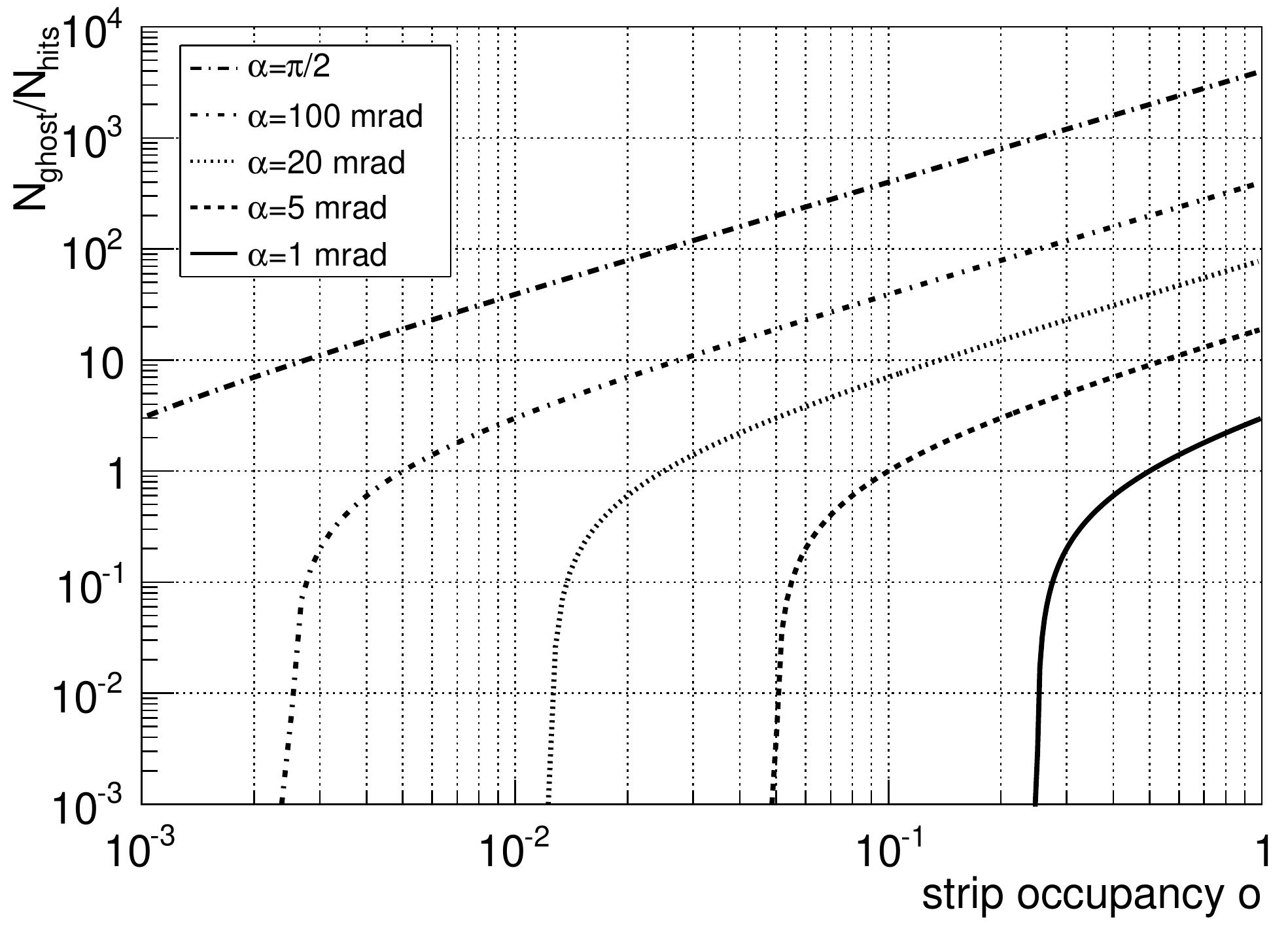}
\caption{The ratio of the number of ghost hits and the number of particles 
traversing the sensor versus strip occupancy. The curves correspond
to five values of the stereo-angle, from 1 mrad to $\pi/2$ rad.}
\label{fig:ghosts}
\end{figure}

Quantitatively, the ghost hit rate depends on the hit density in the detector 
and the detector design (pitch and sensor dimensions). 
In Figure~\ref{fig:ghosts} the number of ghost hits in
 in 100 $\times$ 100 $\mathrm{mm}^2$ sensors with 25 $\mu \mathrm{m}$ pitch is 
compared to the real occupancy due to particles incident on the sensor.
The five curves correspond to stereo angles of from 1 milliradian to 90$^\circ$.
For a given stereo angle the ghost hit rate is negligible until a critical
occupancy is reached. Beyond that point the rate increases very rapidly.
Soon ghost hits are more abundant than the correct combinations and form
an important source of confusion to the pattern recognition stage.
The critical occupancy depends strongly on the stereo angle; smaller stereo 
angles help delay the onset of sufficient ghost
hit contributions to large real occupancies.
In practice, this  feature of micro-strip detectors limits their use to 
environments where the occupancy is below the percent level. For the innermost 
disks in the LC detectors the peak occupancy is well above this level, 
adding yet another argument for the preference of pixelated devices in the 
innermost disks.

\subsection{Precision of the $r$-measurement}
\label{sec:strips}

The $r\phi$ measurement is crucial for the determination of the particle 
momentum. The innermost layers, that play an important role in the
reconstruction of the primary and secondary vertices, must measure a second 
coordinate to constrain the longitudinal impact parameter and polar angle
of the trajectory. The measurement of the $r$-coordinate in the outermost 
disks, however, does not affect the final trajectory parameters significantly.
The $r$-coordinate is measured to yield sufficient constraints on the 
trajectory for pattern recognition to converge. 

In the previous section we have adopted a small stereo angle to avoid
the presence of ghost hits. The combination of the two stereo measurements 
allows to reconstruct 2-dimensional space points, where the resolution
on the $r\phi$ and $r$ coordinates is given\footnote{It is easy to see 
that these relations hold for the 
binary case, where $\sigma = p/\sqrt{12}$, but they hold quite generally. 
Consider two strips with pitch $p$ that cross under an angle $\alpha$. The 
intersection forms a rhombus with short axis $w = p/\cos{(\alpha/2)}$ and 
long axis $h = p/\sin{(\alpha/2)}$. The projection along these axes yields 
triangular distributions with $RMS = w/\sqrt{24}$ and $RMS = h/\sqrt{24}$, 
hence the factor $\sqrt{2}$ in the denominator.} in terms of the space point 
resolution $\sigma$ of each of the measurements and the stereo angle 
$\alpha$:
\begin{flalign*}
\begin{array}{@{}>{\displaystyle}l@{}>{\displaystyle{}}l@{}}
\hskip 5cm \sigma (r \phi) = \frac{\sigma}{\sqrt{2}  \cos{(\alpha/2)}}, \\
\hskip 5cm \sigma (r) = \frac{\sigma}{\sqrt{2}  \sin{(\alpha/2)}}
\end{array}
&&
\label{equation:resolution}
\end{flalign*}
For small values of $\alpha$ the
$r \phi$ measurement improves by a factor of approximately $1/\sqrt{2}$. 
The combination of both 1D measurements yields an $r$ resolution of 
approximately $\frac{\sigma}{2 \alpha}$. For $\alpha = $ 100 mrad, the $r$ resolution
is 20 $\sigma$, i.e. approximately 100 $\mu \mathrm{m}$ in typical detectors. 

During the track finding phase a seed is grown into a full-length
track by extrapolation to the following layer and addition of compatible
hits. A poor measurement of the $r$-coordinate render the extrapolation 
to the following layer inaccurate. If the uncertainty on the extrapolated 
position increases too much, many spurious hits may be found compatible with the
poorly consrained trajectory of the track stub. 

To quantify this argument we propagate seeds generated in either highly granular
or low-density regions\footnote{The former option corresponds to {\em inside-out}
track reconstruction, starting from the vertex detector and pixel disks,
while {\em outside-in} reconstruction starts from the outer silicon micro-strip
layers located in the region with lower hit density} with a rudimentary implementation of the combinatorial algorithm 
used in the LHC experiments. Each seed is propagated to the next layer
with the Kalman filter also used for the final track fit. If a hit is
found whose location is compatible with the extrapolated position within
error, the hit is added to the track stub and propagated to the next
layer. If no compatible hits are encountered, the candidate is not propagated
further\footnote{In real-life implementations a limited number of missing 
hits is allowed to account for inefficiency of the detector layers. Here,
no inefficiencies are simulated and a hit must be found on each layer that
is crossed.}. If multiple hits are found, a new track candidate is spawned
for each compatible hit, and all track candidates are propagated 
in parallel (hence the name
combinatorial). If sufficient constraints are available,
the {\em fake} candidates encounter no compatible hits on further layers
and do not grow into full-length tracks. 
However, if the probability to add spurious hits is non-negligible in all 
layers, or the number of layers is limited, the resulting confusion may 
not be resolved.

\begin{figure}
\centering
\includegraphics[width=0.49\columnwidth]{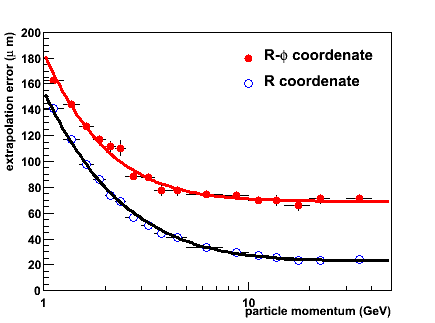}
\includegraphics[width=0.49\columnwidth]{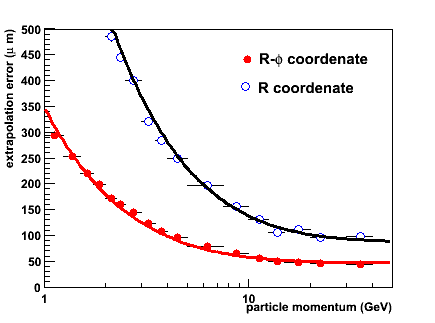}
\caption{The uncertainty on the extrapolated $r \phi$- (closed markers) and $r$-coordinate (open markers). The leftmost panel corresponds to the extrapolation of a pixel triplet to the fourth disk. The rightmost panel corresponds to the extrapolation from the 5th to the 6th disk.}
\label{fig:extrauncertainty}
\end{figure}

As an example, we consider inside-out track finding in three pixel 
disks and four micro-strip disks. The distance 
between disk 1 and 2, 2 and 3, and 3 and 4 is 11 cm. From there on, the 
distance is increased to 25 cm. All layers measure $r \phi$ with
10 $\mu \mathrm{m}$ precision. The pixel disks achieve the same precision in
$r$, while the micro-strip disk provide no constraints on $r$. 
The uncertainty in the $r \phi$- and $r$-coordinate is shown in 
Figure~\ref{fig:extrauncertainty} as a function of particle momentum for
two different steps in the inside-out pattern recognition process. 
The leftmost panel corresponds to the extrapolation over 11 cm 
of a triplet of pixel detector hits to the fourth tracking disk. Three precise measurements
of the $r$-coordinate are sufficient to constrain the trajectory in
the $rz$ projection (where the track model reduces to a straight line in the
high momentum limit). The three equally precise $r \phi$ measurements
provide a poor constraint of the track curvature, due to the short
lever arm of the first disks. Therefore, the $r \phi$ measurement
is predicted with relatively poor precision. The rightmost panel
shows the situation after the track candidates are propagated 
through another couple of disks equipped with micro-strip detectors.
To clearly show the difference, these detectors are assumed to 
measure only the $r \phi$-coordinate. As a consequence of the addition
of two further points with precise $r \phi$ measurements and large
lever arm the $r \phi$-prediction is improved considerably for high
momentum tracks. At low momentum the large separation of the disks
degrades the precision considerably. The $r$-measurement, on the
other hand, is degraded very strongly. The lack of further measurements
to constrain the relevant degrees of freedom of the track fit leads
to a steady growth of the uncertainty of the extrapolation.

The precision with which track seed and stubs can be extrapolated to the
next layer is crucial for pattern recognition. If the
uncertainty in the extrapolated position can be kept to a minimum, 
few spurious hits are compatible (within errors) with the trajectory. 
A quantitative marker is the area of the error ellipse in each 
extrapolation step. The axes of the ellipse are given by the uncertainties 
in $r$ and $r \phi$ position of the extrapolated track. Any hit in the
area of the ellipse must be assigned to the track.

\begin{figure}
\centering
\includegraphics[width=\columnwidth]{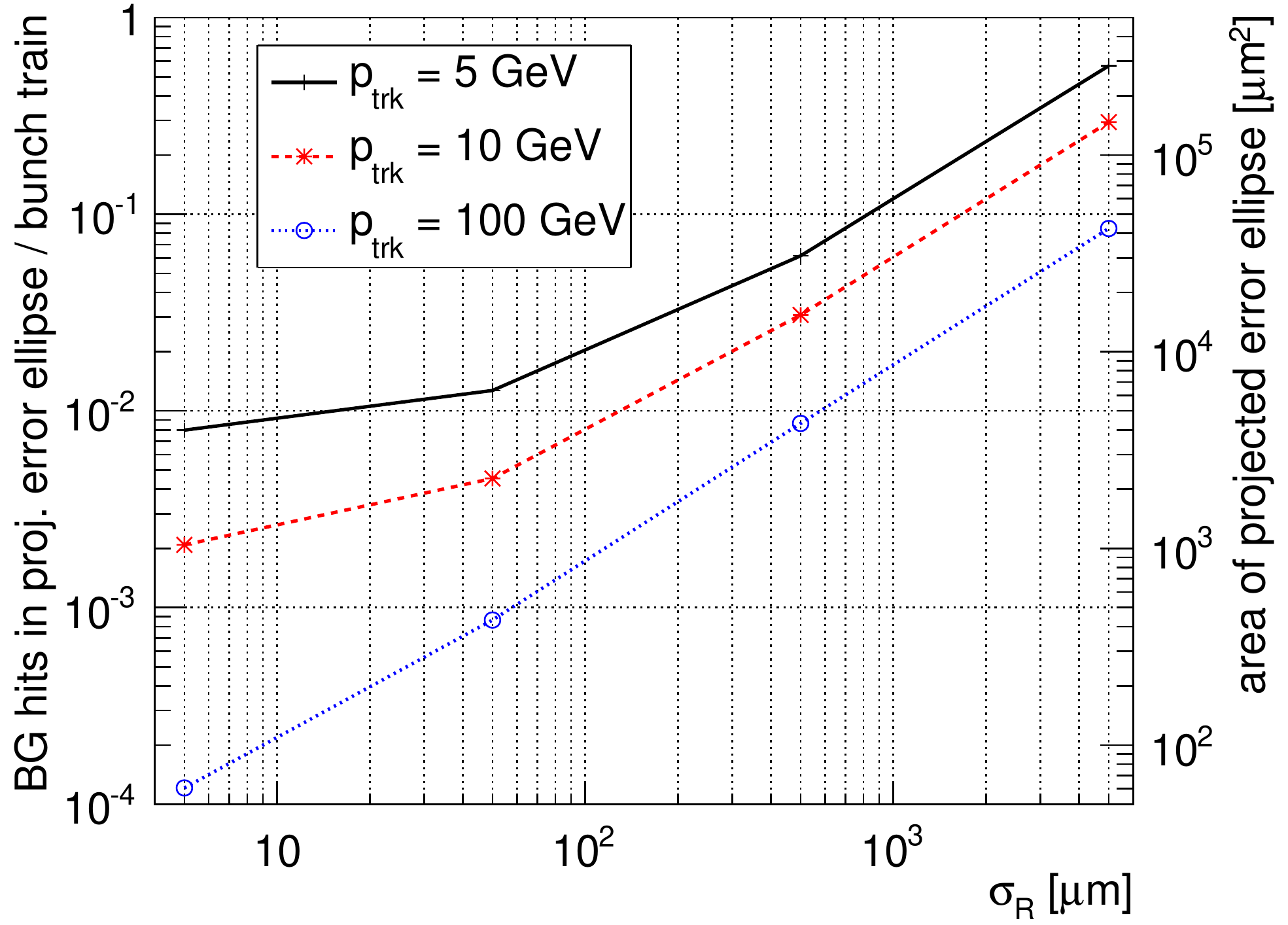}
\caption{Expected background hit rates inside track-extrapolation error ellipses projected from a track stub formed by measurements on six $\mu$-strip disks onto the outermost forward pixel layers. The $r\phi$-resolution is fixed at 5 $\mu\mathrm{m}$ and the resolution in $r$ is varied between 5 $\mu\mathrm{m}$ and 5 mm. The scale on the right axis gives the are of the respective error ellipses. The background occupancy is assumed to be 2 hits per $\mathrm{mm}^2$ and bunch train.  }
\label{fig:error_ellipse_vs_rresolution}
\end{figure}

In Reference~\cite{lcd-note-2011-031} the area of the error ellipse in 
outside-in track finding 
is evaluated for different assumptions on the precision of the 
$r$-measurements of the forward tracking system.
In Figure~\ref{fig:error_ellipse_vs_rresolution}, the area
of the error ellipse on disk $N$ is shown.  
The x-axis ranges from an $r$-resolution below 10 $\mu \mathrm{m}$, 
corresponding to highly granular devices with rectangular pixels, to the 
poor $r$-measurement of micro-strip detectors with a small stereo angle. 
The intermediate region corresponds to elongated pixels with an $r$-dimension 
of several 100 $\mu \mathrm{m}$. 
A second scale indicates how many background hits are 
typically found within this area, which is a good measure of the confusion term
in each pattern recognition step.
The result for high momentum tracks is indicated by the (blue) dotted curve. 
In this case the area depends linearly on the $r$-resolution.
For low-momentum tracks the confusion increases. A clear saturation is moreover
observed, as at one point the multiple scattering contribution to the 
uncertainty prevents further improvement of the extrapolated position.
This result shows that a precise $r$-measurement can render pattern 
recognition much more robust. Beyond an $r$-precision of the
order of 100 $\mu \mathrm{m}$ little is gained for low momentum
tracks, that are the hardest to reconstruct efficiently and cleanly. 
We conclude, therefore, that the design of the forward region should aim for
moderately precise $r$-measurements in all tracking layers.

A number of stress tests have been 
performed to evaluate how strongly the standalone pattern recognition 
performance of the Forward Tracking Disks depends on the design 
assumptions. A strong increase in the material or in the 
read-out time can degrade the pattern recognition performance to an 
unacceptable level. 

In the forward tracking system, pattern recognition must converge after only
a small number of measurements separated by large distances. The devices that
equip the Forward Tracking Disks should provide two-dimensional space points
to reduce the ambiguities in the track finding stage. We find that the 
$r$ measurement with moderate precision that is readily provided by micro-strip
detectors with a small stereo angle is adequate for this purpose.

\section{Summary \& conclusions}
\label{sec:discussion}

In a previous paper~\cite{forwardnote} we have argued that as $e^+ e^-$ 
colliders are built to reach higher and higher center-of-mass energy, 
the relevance of the forward and backward regions of the detector 
design for the overall performance of the experiment increases considerably. 

In this work we identify the main challenges of the forward tracking system
and relate them to the detector design in a quantitative fashion.

The transverse momentum resolution is degraded in the forward 
region, approximately as $1/\sin^{3/2}{\theta}$, due to the unfavourable 
orientation of the magnetic field. To maintain good peformance for tracks 
emitted at small polar angle, emphasis should be given to the development 
of detectors with excellent spatial resolution for the $r\phi$-coordinate, 
while maintaining the strict material budget.

The vertex reconstruction performance for
charged particles emitted at small polar angle is also degraded.
An endcap detector equipped with precise and thin pixel detectors can check
the $1/\sin{\theta}^{3/2}$ growth of the impact parameter resolution, but
this requires a very strict control of the material in the beam pipe 
and the services and support in the barrel-endcap transition. 

Efficient and clean track reconstruction demands highly granular devices in 
the innermost regions of the detector, where the background density is highest. 
Robust pattern recognition moreover requires the determination with moderate
precision of the $r$-coordinate of hits in all forward tracking layers.

\section{Acknowledgements}

This work has been partly supported by Grants FPA2010-22163-C02-01 and FPA2010-21549-C04-4 of the Spanish Ministry of Economy and Competitiveness; and  by the European Commission within Framework Programme 7 Capacities, Grant Agreement 262025 (AIDA). We would like to acknowledge the ILD software group that developed many of the tools used to obtain the reported results. M. Vos would moreover like to thank the LCD group at CERN for their hospitality.

\begin{footnotesize}
\bibliographystyle{JHEP}
\bibliography{forward2}{}
\end{footnotesize}
\end{document}